\documentclass[fleqn,usenatbib]{mnras}

\usepackage{natbib}
\usepackage{color}

\usepackage[T1]{fontenc}

\DeclareRobustCommand{\VAN}[3]{#2}
\let\VANthebibliography\thebibliography
\def\thebibliography{\DeclareRobustCommand{\VAN}[3]{##3}\VANthebibliography}

\usepackage{graphicx}	
\usepackage{amsmath}	
\usepackage{amssymb}	
\usepackage{newtxtext,newtxmath}

\newcommand{\code}[1]{\texttt{#1}}

\title[Bubbles and Cold Fronts]{The interaction between rising bubbles and cold fronts in cool core clusters}

\author[A.C. Fabian et al.]{
A. C. Fabian,$^{1}$\thanks{E-mail: acf@ast.cam.ac.uk}
J. Zuhone,$^{2}$ and 
S. A. Walker,$^{3}$ 
\\
$^{1}$Institute of Astronomy. Madingley Road, Cambridge  CB3 0HA, UK\\
$^{2}$Center for Astrophysics $\vert$~Harvard~\&~Smithsonian, Cambridge, MA 02138, USA\\
$^{3}$Department of Physics and Astronomy, The University of Alabama in Huntsville, Huntsville, AL 35899, USA\\
}

\date{Accepted XXX. Received YYY; in original form ZZZ}

\pubyear{}

\begin{document}
\label{firstpage}
\pagerange{\pageref{firstpage}--\pageref{lastpage}}
\maketitle

\begin{abstract}
We investigate whether the swirling cold front in the core of the Perseus Cluster of galaxies has affected the outer buoyant bubbles that  originated from jets from the  Active Galactic Nucleus in the central galaxy NGC1275. The inner bubbles and the Outer Southern bubble lie along a North-South axis through the nucleus, whereas the Outer Northern bubble appears rotated about 45 deg from that axis. Detailed numerical simulations of the interaction indicates that the Outer Northern bubble may have been pushed clockwise accounting for its current location.  Given the common occurrence of cold fronts in cool core clusters, we raise the possibility that the lack of many clear outer bubbles in such environments may be due to their disruption  by cold fronts.  
\end{abstract}




\section{Introduction}
The Perseus cluster of galaxies has proven to be an excellent demonstration of
the action of Active Galactic Nucleus (AGN) Feedback in clusters, as well as a
testbed for understanding the processes involved. A pair of inner cavities or
bubbles on a North-South axis either side of the AGN in the central galaxy
NGC1275 were first imaged in X-rays by ROSAT \citep{Boehringer1993} and then
studied in detail with Chandra \citep{Fabian2000,Fabian2003,Fabian2006}.  An outer bubble is seen
further along the axis to the South with its partner lying to the North-West at
an angle of about 45 deg from the axis of the inner bubbles. The axis of the
inner bubbles is also marked to the North by optical H$\alpha$ filamentation, also seen
faintly in soft X-rays,  extending 50 kpc from the AGN (see Fig. 1).

Energy from the accreting black hole at the centre of NGC1275 powers jets  seen
seen at radio wavelengths as 3C84 \citep{Pedlar1990,Gendron-Marsolais2020}. The jets inflate the
inner bubbles with plasma composed of relativistic particles and magnetic field which are  buoyant in the denser ICM and  rise to form a new set of outer
bubbles.  The power of the bubbling process of about $10^{44.5}$ to
$10^{45}$ erg s$^{-1}$ is comparable to the energy loss of the cool core through
radiating the X-ray emission we see. The whole bubbling process constitutes a
dramatic example of AGN Feedback. 

The  off-axis arrangement of the NW bubble has in the past been considered to be due to
precession of the central jets (\citealt{Dunn2006}; \citealt{Falceta-Goncalves2010}; see also \citealt{SternbergSoker2008}) or jet bending due to the bulk motion of the
intracluster medium (ICM) relative to the AGN \citep{Soker2006}.  Such effects
would be expected to have influenced both of the outer bubbles but only the
Northern one appears to be affected.

Here we consider the effects of the inner cold front that circles the centre of
the cluster core. The front is due to sloshing motions induced in the core by
the close passage of a subcluster merger a few billion  years ago. The NW outer
bubble lies just inside the cold front where streaming motions occur that can
carry the bubble in the required direction. An optical filament  extends beyond
the cold front to the N but then seems to twist also to the W. The S outer bubble
has an unexplained double structure and lies beyond the cold front. We are
of course viewing the system as projected on the plane of the Sky, whereas the
jets that create the structure are  oriented towards us in the N and away in the
S. The inclination to the line of sight of the jets is a matter of
debate and varies widely on which observations, from Radio (65 deg to line of sight) to Gamma-rays (23 deg) are
being considered (see \citealt{Hodgson2021}). 
This uncertainty is  due to which part of the inner jet system dominates
the emission at different wavelengths and is a consequence of the dense clumpy
central environment through which the jets initially propagate. The location of the most
massive clumps could influence the final jet direction in a manner that is
independent of the initial direction and so lead to the Northern Outer Bubble
being offset from the Southern one. This would however require that the anisotropy
lasts for the several Myr needed to form the bubbles, which is many dynamical
times for the inner region and therefore unlikely.

Here we study the possible interaction between the
Northern Outer Bubble and  the Inner Cold Front. The Southern Bubble lies outside the cold front and we
presume that the geometry is such that it did not directly interact with it. We do not aim for a perfect model but just illustrate using simulations
that a twist in the direction of motion of a bubble of the amplitude seen is expected as it
approaches and passes through a front. 

\begin{figure}
	\includegraphics[width=\columnwidth]{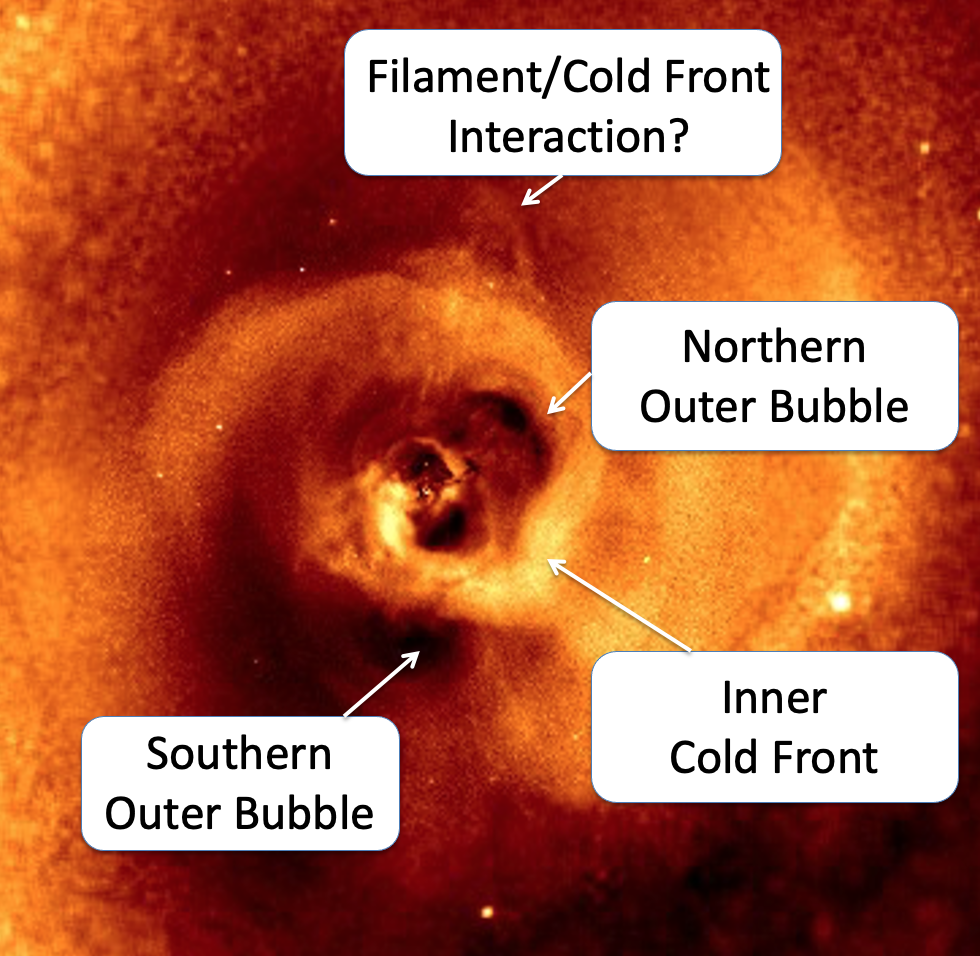}
	\includegraphics[width=\columnwidth]{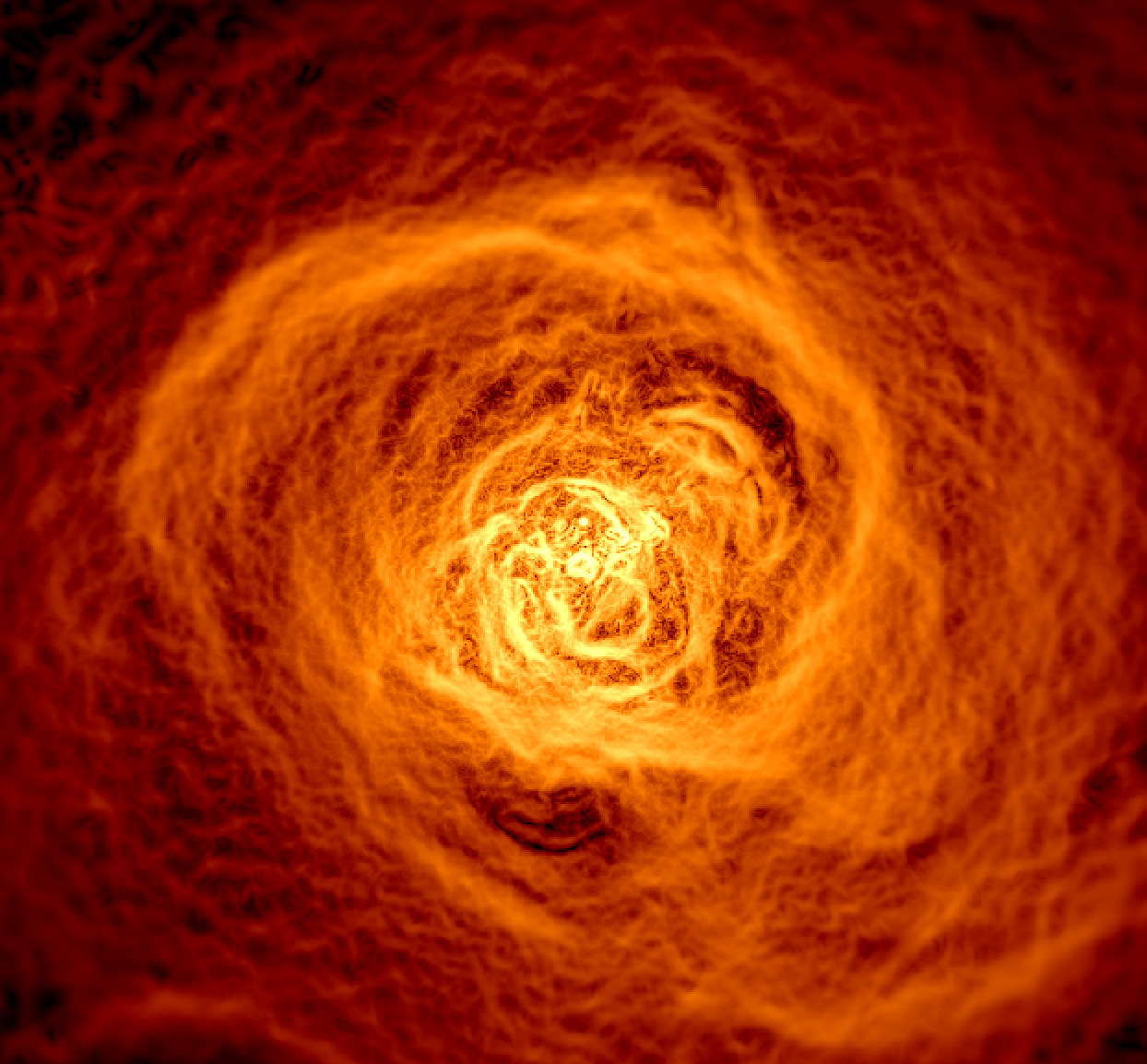}
    \caption{The core of the Perseus cluster imaged in X-rays by Chandra and displayed after application of (Top) an unsharp-mask and (Bottom) a Gaussian Gradient Method, which emphasises edges (from \citealt{Walker2017}).  The upper image is about 400 kpc across and the lower is 200kpc. Salient features are labelled in the upper image andthe next loop of the spiralling cold front is seen near the Western edge.    }
    \label{fig:PerBubbles}
\end{figure}

\begin{figure*}
	\includegraphics[width=\textwidth]{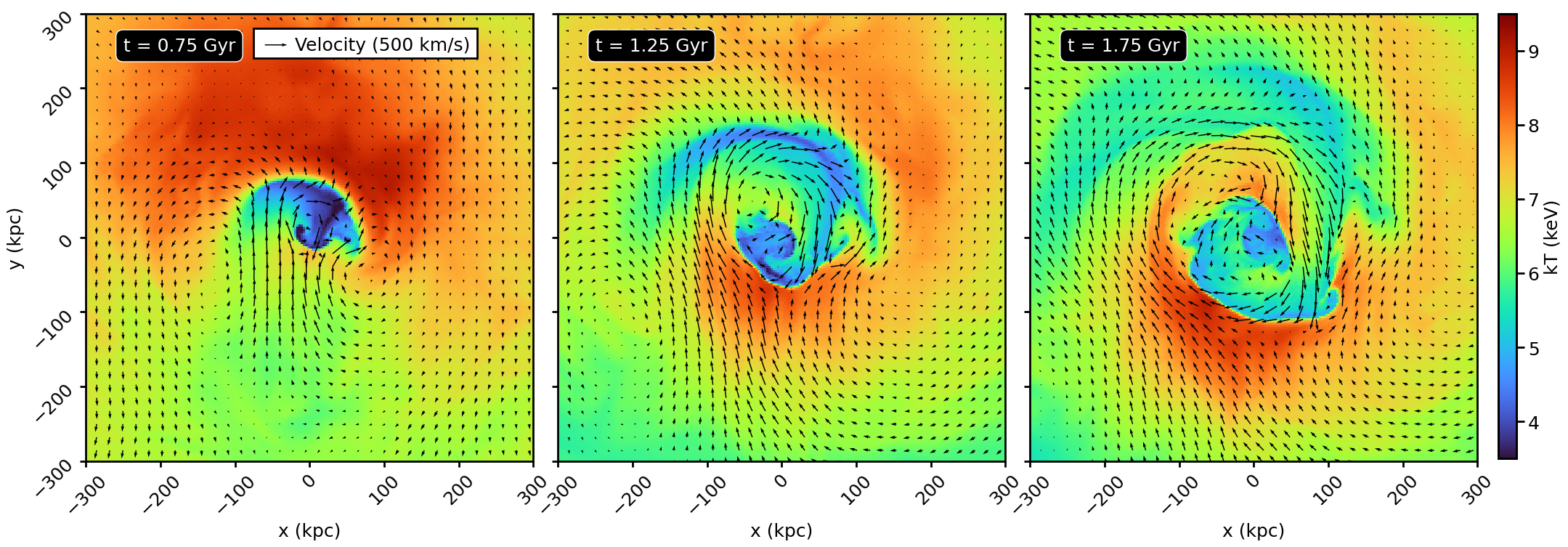}
    \caption{Slices through the gas temperature with velocity vectors overlaid for 
	         the three epochs of the merger simulation used in this work. Each 
			 slice is centered on the cluster potential minimum.}
    \label{fig:sloshing_slices}
\end{figure*}


\begin{figure*}
	\includegraphics[width=\textwidth]{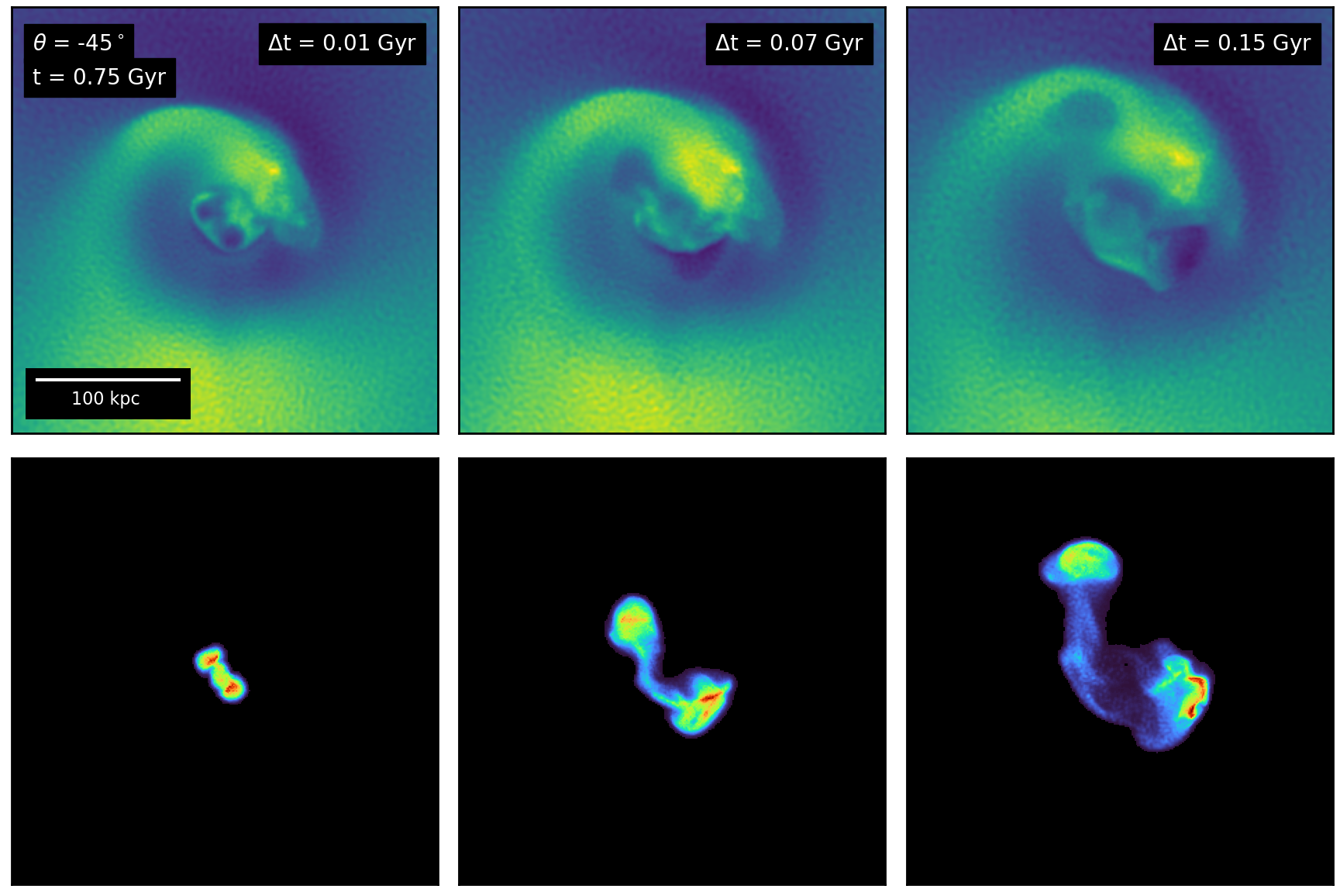}
    \caption{X-ray surface brightness residuals (top panels) and
             projected CR energy density (bottom panels) of our sloshing simulation where the jet was ignited at $t$ = 0.75~Gyr at an angle of $\theta$ = -45$^\circ$ from the $x$-axis, observed at three different epochs.\label{fig:example1}}
\end{figure*}

\begin{figure*}
	\includegraphics[width=\textwidth]{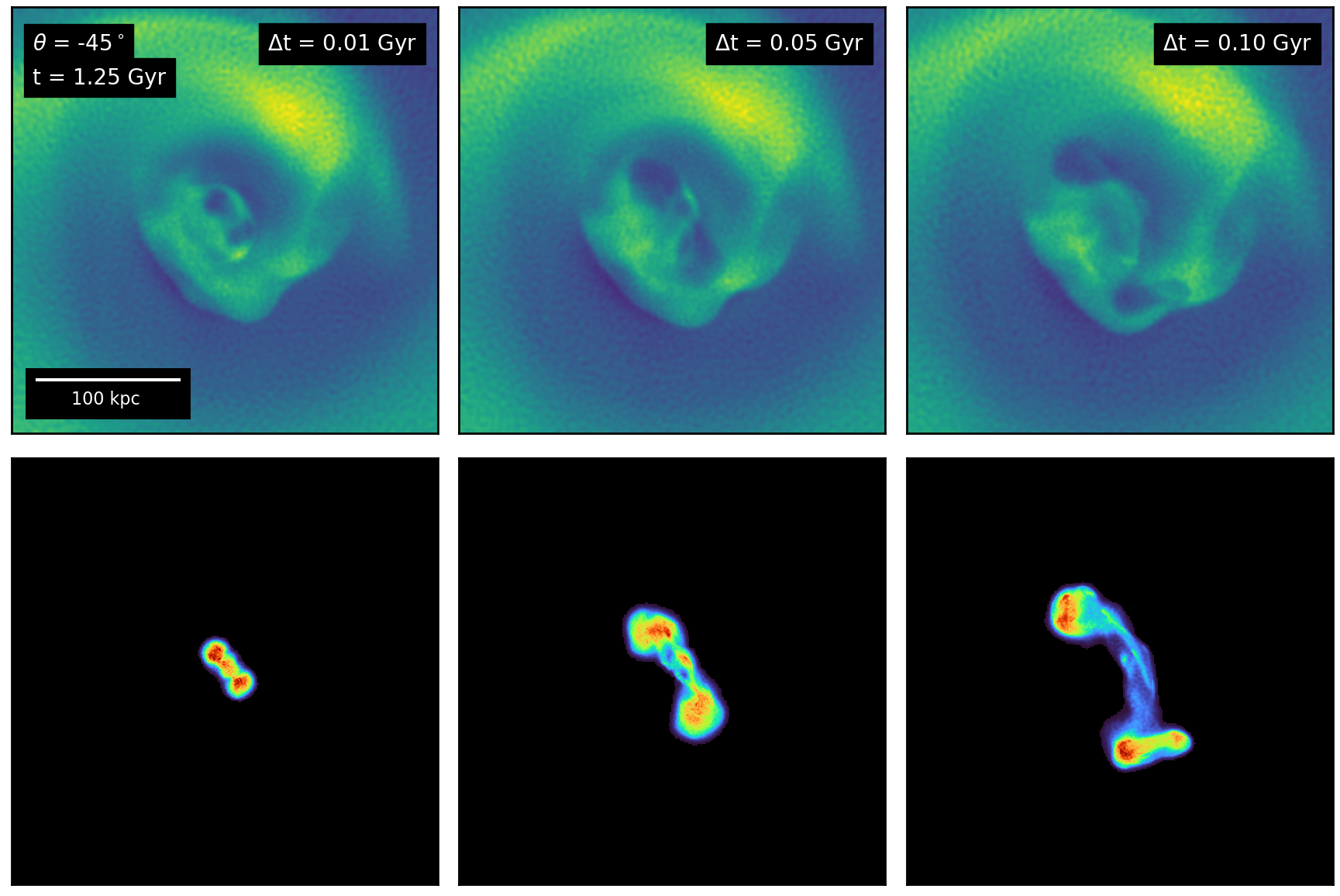}
    \caption{X-ray surface brightness residuals (top panels) and
             projected CR energy density (bottom panels) of our sloshing simulation where the jet was ignited at $t$ = 1.25~Gyr at an angle of $\theta$ = -45$^\circ$ from the $x$-axis, observed at three different epochs.\label{fig:example2}}
\end{figure*}

\begin{figure*}
	\includegraphics[width=\textwidth]{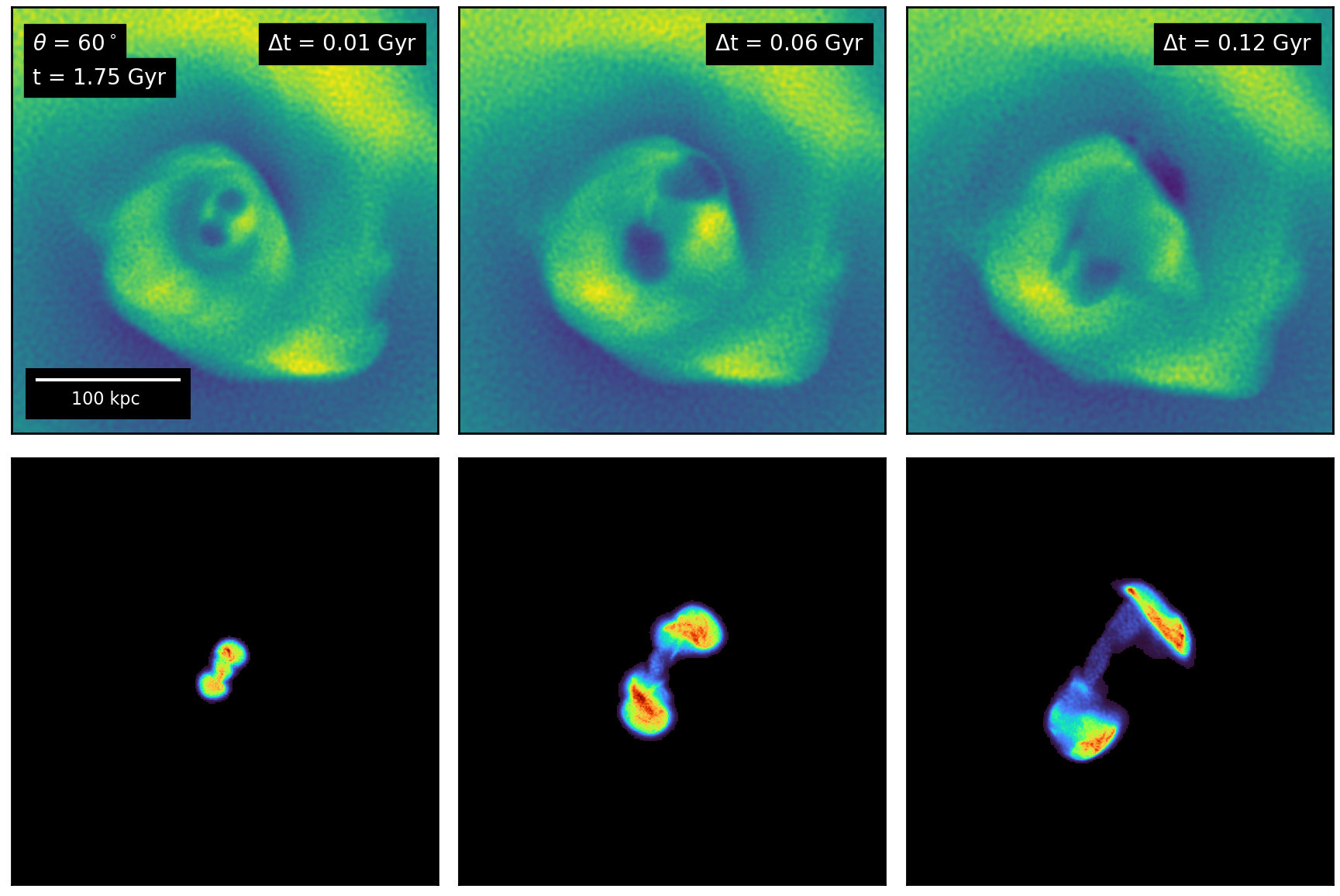}
    \caption{X-ray surface brightness residuals (top panels) and
             projected CR energy density (bottom panels) of our sloshing simulation where the jet was ignited at $t$ = 1.75~Gyr at an angle of $\theta$ = 60$^\circ$ from the $x$-axis, observed at three different epochs.\label{fig:example3}}
\end{figure*}

\section{Brief Description of Sloshing Cold Fronts}

Cold fronts are features in X-ray observations which are characterized by sharp
surface brightness (SB) discontinuities in the ICM in which the measured
temperature is colder on the brighter (and therefore denser) side of the
feature. Simulations indicate that these features can be formed quite easily in
mergers between galaxy clusters. This occurs when dense, low-entropy gas is
brought into contact with more diffuse high-entropy gas and forms a contact
discontinuity which travels at subsonic velocities. A recent review of cold
fronts, including the details of the formation mechanisms in various
circumstances, can be found in \citep{ZR2016}. 

For a massive, cool-core cluster like Perseus, the cold fronts likely formed as
a result of an off-center encounter with a subcluster. This process has been
simulated in a number of works, including \citet{AM06,ZuHone2010,Roediger2011,Roediger2012}. As the subcluster approaches the center of the larger cluster, it
gravitationally accelerates both the gas and DM components of the core. However,
the movement of the gas component is slowed by the ram pressure of the
surrounding medium, and so it becomes spatially separated from the Dark Matter. After the
subcluster has passed and its influence has weakened, the cold, low-entropy gas
of the core falls back into the DM-dominated potential well, and begins to
oscillate around it in a process dubbed ``sloshing,'' which
produces cold fronts. Because the encounter with the subcluster was off-axis, it
transfers angular momentum to the gas, and the sloshing gas motions and the cold
fronts they produce are spiral-shaped. The cold fronts move radially outward
slowly, with Mach number $\mathcal{M} \sim 0.1$, but the tangential flow underneath the front surface on the denser side
is much faster, with $\mathcal{M} \sim 0.3-0.5$. 

Figure \ref{fig:sloshing_slices} shows slices of the gas temperature for three
different epochs in a simulation of gas sloshing that will be described in
Section \ref{sec:sim_methods}. The sharp gradients in temperature indicate the
location of the cold fronts. Velocity vectors are overlaid, indicating that the
flow pattern is largely tangential to the front surfaces and underneath them. As
we show in Section \ref{sec:sim_results}, these tangential motions underneath
the fronts can have significant effects on the evolution of cavities produced by
the AGN at the core. 

\section{Results}

\subsection{Simulation Methods}\label{sec:sim_methods}

To carry out our simulations of gas sloshing combined with AGN jets, we use the
\code{AREPO} code \citep{Springel2010}, which employs a finite-volume Godunov
method on an unstructured moving Voronoi mesh to evolve the equations of
magnetohydrodynamics (MHD), and a Tree-PM solver to compute the self-gravity
from gas and dark matter. The magnetic fields are evolved on the moving mesh
using the Powell 8-wave scheme with divergence cleaning employed in
\citep{Pakmor2013} and in the IllustrisTNG simulations \citep{Marinacci2018}.
These simulations also include dark matter particles, which make up the bulk of
the cluster's mass and only interact with each other and the gas via gravity. 

The simulations model a large, Perseus-like, cool-core cluster with $M \sim
10^{15}~M_\odot$, with initial conditions taken from previous works
\citep{AM06,ZuHone2010,ZuHone2016,ZuHone2018,ZuHone2019,ZuHone2020}. This
cluster has an off-axis encounter with a DM-only subcluster of $M \sim 2 \times
10^{14}~M_\odot$, which initiates sloshing motions which produce cold fronts.
The ICM in the simulation has an initially tangled magnetic field with an
initial plasma parameter $\beta = p_{\rm th}/p_B = 100$. We use in this work the
version of this simulation from \citet[][hereafter Z20]{ZuHone2020}, with the
same number of particles and cells, and mass and spatial resolution. The merger
takes place within the $x-y$ plane, so that most (but not all) of the gas
motions are in these directions. More details of the setup of this merger
simulation are given in Section 2 of that paper. 

We also use the same jet model as in Z20, originally presented in
\citet{Weinberger2017}. This method injects a bi-directional jet which is
kinetically dominated, low density, and collimated. Kinetic, thermal, and
magnetic energy is injected into two small spherical regions a few kpc from the
location of a black hole particle. The material injected by the jet is marked by
a passive tracer field $\rho_{\rm jet}$ and is advected along with the fluid for the rest of the simulation.
In contrast to Z20, the material injected by the jet is composed of a 50/50 mixture of two fluids, one with $\gamma = 4/3$ representing cosmic rays (hereafter CRs) and another with $\gamma = 5/3$, the same as the ICM.

A black hole particle is placed at the cluster potential minimum, which serves
as the site of the jet injection by the AGN. In each simulation, the jets are
fired with a power $P_{\rm jet} = 10^{45}$~erg~s$^{-1}$ for a duration of
$t_{\rm jet} = 50$~Myr, so the resulting total energy injected is $E_{\rm jet}
\sim 1.6 \times 10^{60}$~erg in each direction, which is a sum of kinetic,
thermal, and magnetic energy. In the jet region, the magnetic and thermal
pressures are equal ($\beta_{\rm jet} = 1$), and the injected magnetic field is
purely toroidal.

\subsection{Simulation Results}\label{sec:sim_results}

We begin by choosing epochs of the merger simulation shortly after the
subcluster passage but long enough for cold fronts to have developed. This
ensures that the gas motions are fast and the first cold fronts to form have not
yet propagated to a large radius. In reality, since structure formation is a
continuous process, encounters with small subclusters may be frequent enough
that new sloshing motions and cold fronts may be generated fairly frequently, so
our choice of ``favorable'' epochs to affect the evolution of AGN bubbles is not
arbitrary. 

We examine three epochs, $t = 0.75$, 1.25, and 1.75~Gyr, where the time is
measured after the core passage of the subcluster (these are the three epochs
shown in Figure \ref{fig:sloshing_slices}). For each epoch, we perform several
simulations, varying the jet orientation parameterized by the angle $\theta$,
defined as the angle measured counterclockwise from the $x$-axis. We choose
$\theta \in [-60^{\circ}, -45^{\circ}, -30^{\circ}, 0^{\circ}, 30^{\circ},
45^{\circ}, 60^{\circ}, 90^{\circ}]$ (other orientations being degenerate with
these due to the bipolar nature of the outflow).   

Figures \ref{fig:example1}-\ref{fig:example3} shows the projected X-ray SB
residuals and the projected CR energy density for a few representative cases
from our simulations. The bubbles are visible in as cavities in the X-ray SB
maps and as concentrations of CR energy density. Each case clearly shows that
the bubbles begin along the jet axis, but as they rise they encounter the
fast-moving ICM under the cold fronts and are subsequently blown off this axis. Fig. \ref{fig:example1} shows rotation of the upper bubble similar to a similar degree  seen in the Perseus cluster. 
Additionally, the gas motions stretch the bubbles in the tangential direction,
producing pancake-shaped cavities. Both of these features of the cavities are
seen in Perseus (Figure \ref{fig:PerBubbles}). The simulations suggest that the Southern bubble is unlikely to have escaped unscathed. The bending of the jets provided
by the sloshing motions also produces features in the CR energy density maps
that resemble wide-angle tails (WATs), which have already been shown to be
easily produced in simulations by subsonic gas motions such as sloshing \citep{Loken1995,Mendygral2012}. We note that radiation losses on the most energetic electrons (spectral ageing) are responsible for the lack of detectable radio emission at 1 GHz from the cosmic rays in the outer bubbles of Perseus.   

For completeness, we also show similar maps of all jet orientations for all
three epochs in Appendix \ref{sec:supp_plots}. These plots show other
interesting features depending on the orientation, including bubbles which split
into two, resulting in three observed cavities, and WATs of various shapes at
late stages.

\section{Discussion}
We have shown that that interactions between outer bubbles and a sloshing cold front can deflect the trajectories of the bubbles and cause their positions to depart from the line of their initial trajectory. Also, passage through a cold front can seriously distort and possibly destroy a bubble. With that in mind we have searched the literature for evidence of multiple bubbles. Large compilations of bubbles (or X-ray cavities)  appear in \citet{Diehl2008} and \citet{Birzan2020}, and there are many reports on individual sources. One striking result is that multiple pairs of bubbles in the same system are rare.

Clear examples are found in some groups of galaxies, such as  NGC 5813 \citep{Randall2011,Randall2015} and NGC 5044 \citep{Schellenberger2021}. The bubbles are smaller than those in clusters and groups tend not to have cold fronts, contrary to the cool cores of rich clusters where perhaps two-thirds have cold fronts \citep{Ghizzardi2010,ZuHone2016}. Hydra-A has multiple sets of relatively large bubbles at the centre of a cluster \citep{Wise2007} which does not appear to host a cold front but a very large shock front at a radius of 200--300 kpc \citep{Nulsen2005}.  A precessing jet model for the Hydra-A jets has been presented by \citet{Nawaz2016}. As mentioned above, most cool-core clusters with bubbles show just one pair\footnote{The nearby M87 system shows several small bubbles adjacent to the E inner bubble \citep{Forman2007}. These are unlikely to be related to the interactions discussed here.}. It is of course possible that outer bubbles are present but undetectable due to reasons of contrast or lower surface brightness (see \citealt{Panagoulia2014} for a discussion of bubble detectability in nearby systems), but none are seen even in objects with deep X-ray images.

A more likely explanation from the present work is that buoyant outer bubbles may be sufficiently distorted or destroyed by interacting with cold fronts. This would mean that the effects like the interaction seen in the Perseus Cluster may be common. For example, \citet{Sanders2009} presented results on the cluster 2A 0335+096, which shows evidence of both sloshing cold fronts and 4 X-ray cavities which may have been transported tangentially by the sloshing motions or even split apart by them (see the $\theta = 45{^\circ}$, $\Delta{t} = 0.1$~Gyr panel of Figures \ref{fig:sx_260_1} and \ref{fig:cr_260_1} for examples from our simulations in the Appendix). 

One feature that we defer to later work is possible interactions between the H$\alpha$ filaments of the Perseus Cluster and the sloshing cold fronts. The filaments contain both atomic and molecular gas \citep{Hatch2005,Salome2006,Lim2012}. The NW outer bubble in the Perseus Cluster does have a "horseshoe" filament below it \citep{Conselice2001,Fabian2003_Halpha} which may  reveal the fluid flow. The interaction of dense filaments with the swirling lighter gas is not straightforward and will depend on the magnetic coupling between the two media.

There are a number of areas for future work. Our simulated cluster, while a
close match to Perseus in terms of mass, temperature, and shapes of the cold
fronts, is not an exact match, especially with respect to the effects of AGN
feedback. In subsequent works, we will attempt to simulate a closer match to the
Perseus Cluster, reproducing the sizes and orientation of the cold fronts and
cavities, as well as the properties of the velocity field already observed by
\textit{Hitomi} \citep{Hitomi2016}. Indeed the highest velocity sector reported from the analysis by \citep{Hitomi2018} is the one covering the Northern Outer Bubble and neighbouring cold front, the site of the interaction at the heart of this paper. Velocity fields in the cores of the Perseus and Virgo clusters have been obtained at  lower spectral resolution with \textit{XMM-Newton}  by (\citealt{Sanders2020}; Gatuzz et al. 2021, submitted). In the near future  \textit{XRISM} \citep{XRISM2018} will have a detector similar to that on Hitomi (with a resolution of 5 eV) and in the early 2030s \textit{Athena} \citep{Barcons2017} will observe clusters with yet higher spectral resolution (2.5 eV) and good spatial resolution (about 6 arcsec). A deeper Chandra image of the Perseus cluster would reveal the interaction and properties of the outer bubbles and inner cold front in greater detail. 

In summary, the offset location of the Outer Northern Bubble in the Perseus Cluster suggests that a strong interaction is taking place with the Inner Cold Front. Our simulations coupled with the lack of multiple bubble systems in  rich clusters imply that cold fronts disrupt outer bubbles. 

\section*{Acknowledgements}

JAZ acknowledges support from the Chandra X-ray Center, 
which is operated by the Smithsonian Astrophysical Observatory for and on behalf of NASA under contract NAS8-03060. The simulations were performed on the ``Pleiades'' high-performance computing system at the NASA Advanced Supercomputing facility at NASA/Ames Research Center. 

\section*{Data Availability}

The \textit{Chandra} data used in the paper are publicly available from the Chandra Data Archive (CDA). FITS images from the simulation data will be made available on the Galaxy Cluster Merger Catalog at \url{http://gcmc.hub.yt}.



\bibliographystyle{mnras}
\bibliography{bubblesandfronts} 

\begin{thebibliography}{}
\makeatletter
\relax
\def\mn@urlcharsother{\let\do\@makeother \do\$\do\&\do\#\do\^\do\_\do\%\do\~}
\def\mn@doi{\begingroup\mn@urlcharsother \@ifnextchar [ {\mn@doi@}
  {\mn@doi@[]}}
\def\mn@doi@[#1]#2{\def\@tempa{#1}\ifx\@tempa\@empty \href
  {http://dx.doi.org/#2} {doi:#2}\else \href {http://dx.doi.org/#2} {#1}\fi
  \endgroup}
\def\mn@eprint#1#2{\mn@eprint@#1:#2::\@nil}
\def\mn@eprint@arXiv#1{\href {http://arxiv.org/abs/#1} {{\tt arXiv:#1}}}
\def\mn@eprint@dblp#1{\href {http://dblp.uni-trier.de/rec/bibtex/#1.xml}
  {dblp:#1}}
\def\mn@eprint@#1:#2:#3:#4\@nil{\def\@tempa {#1}\def\@tempb {#2}\def\@tempc
  {#3}\ifx \@tempc \@empty \let \@tempc \@tempb \let \@tempb \@tempa \fi \ifx
  \@tempb \@empty \def\@tempb {arXiv}\fi \@ifundefined
  {mn@eprint@\@tempb}{\@tempb:\@tempc}{\expandafter \expandafter \csname
  mn@eprint@\@tempb\endcsname \expandafter{\@tempc}}}

\bibitem[\protect\citeauthoryear{{Ascasibar} \& {Markevitch}}{{Ascasibar} \&
  {Markevitch}}{2006}]{AM06}
{Ascasibar} Y.,  {Markevitch} M.,  2006, \mn@doi [\apj] {10.1086/506508}, \href
  {https://ui.adsabs.harvard.edu/abs/2006ApJ...650..102A} {650, 102}

\bibitem[\protect\citeauthoryear{{Barcons} et~al.,}{{Barcons}
  et~al.}{2017}]{Barcons2017}
{Barcons} X.,  et~al., 2017, \mn@doi [Astronomische Nachrichten]
  {10.1002/asna.201713323}, \href
  {https://ui.adsabs.harvard.edu/abs/2017AN....338..153B} {338, 153}

\bibitem[\protect\citeauthoryear{{B{\^\i}rzan} et~al.,}{{B{\^\i}rzan}
  et~al.}{2020}]{Birzan2020}
{B{\^\i}rzan} L.,  et~al., 2020, \mn@doi [\mnras] {10.1093/mnras/staa1594},
  \href {https://ui.adsabs.harvard.edu/abs/2020MNRAS.496.2613B} {496, 2613}

\bibitem[\protect\citeauthoryear{{Boehringer}, {Voges}, {Fabian}, {Edge}  \&
  {Neumann}}{{Boehringer} et~al.}{1993}]{Boehringer1993}
{Boehringer} H.,  {Voges} W.,  {Fabian} A.~C.,  {Edge} A.~C.,   {Neumann}
  D.~M.,  1993, \mn@doi [\mnras] {10.1093/mnras/264.1.L25}, \href
  {https://ui.adsabs.harvard.edu/abs/1993MNRAS.264L..25B} {264, L25}

\bibitem[\protect\citeauthoryear{{Conselice}, {Gallagher}  \&
  {Wyse}}{{Conselice} et~al.}{2001}]{Conselice2001}
{Conselice} C.~J.,  {Gallagher} John~S. I.,   {Wyse} R. F.~G.,  2001, \mn@doi
  [\aj] {10.1086/323534}, \href
  {https://ui.adsabs.harvard.edu/abs/2001AJ....122.2281C} {122, 2281}

\bibitem[\protect\citeauthoryear{{Diehl}, {Li}, {Fryer}  \& {Rafferty}}{{Diehl}
  et~al.}{2008}]{Diehl2008}
{Diehl} S.,  {Li} H.,  {Fryer} C.~L.,   {Rafferty} D.,  2008, \mn@doi [\apj]
  {10.1086/591310}, \href
  {https://ui.adsabs.harvard.edu/abs/2008ApJ...687..173D} {687, 173}

\bibitem[\protect\citeauthoryear{{Dunn}, {Fabian}  \& {Sanders}}{{Dunn}
  et~al.}{2006}]{Dunn2006}
{Dunn} R.~J.~H.,  {Fabian} A.~C.,   {Sanders} J.~S.,  2006, \mn@doi [\mnras]
  {10.1111/j.1365-2966.2005.09928.x}, \href
  {https://ui.adsabs.harvard.edu/abs/2006MNRAS.366..758D} {366, 758}

\bibitem[\protect\citeauthoryear{{Fabian} et~al.,}{{Fabian}
  et~al.}{2000}]{Fabian2000}
{Fabian} A.~C.,  et~al., 2000, \mn@doi [\mnras]
  {10.1046/j.1365-8711.2000.03904.x}, \href
  {https://ui.adsabs.harvard.edu/abs/2000MNRAS.318L..65F} {318, L65}

\bibitem[\protect\citeauthoryear{{Fabian}, {Sanders}, {Allen}, {Crawford},
  {Iwasawa}, {Johnstone}, {Schmidt}  \& {Taylor}}{{Fabian}
  et~al.}{2003a}]{Fabian2003}
{Fabian} A.~C.,  {Sanders} J.~S.,  {Allen} S.~W.,  {Crawford} C.~S.,  {Iwasawa}
  K.,  {Johnstone} R.~M.,  {Schmidt} R.~W.,   {Taylor} G.~B.,  2003a, \mn@doi
  [\mnras] {10.1046/j.1365-8711.2003.06902.x}, \href
  {https://ui.adsabs.harvard.edu/abs/2003MNRAS.344L..43F} {344, L43}

\bibitem[\protect\citeauthoryear{{Fabian}, {Sanders}, {Crawford}, {Conselice},
  {Gallagher}  \& {Wyse}}{{Fabian} et~al.}{2003b}]{Fabian2003_Halpha}
{Fabian} A.~C.,  {Sanders} J.~S.,  {Crawford} C.~S.,  {Conselice} C.~J.,
  {Gallagher} J.~S.,   {Wyse} R.~F.~G.,  2003b, \mn@doi [\mnras]
  {10.1046/j.1365-8711.2003.06856.x}, \href
  {https://ui.adsabs.harvard.edu/abs/2003MNRAS.344L..48F} {344, L48}

\bibitem[\protect\citeauthoryear{{Fabian}, {Sanders}, {Taylor}, {Allen},
  {Crawford}, {Johnstone}  \& {Iwasawa}}{{Fabian} et~al.}{2006}]{Fabian2006}
{Fabian} A.~C.,  {Sanders} J.~S.,  {Taylor} G.~B.,  {Allen} S.~W.,  {Crawford}
  C.~S.,  {Johnstone} R.~M.,   {Iwasawa} K.,  2006, \mn@doi [\mnras]
  {10.1111/j.1365-2966.2005.09896.x}, \href
  {https://ui.adsabs.harvard.edu/abs/2006MNRAS.366..417F} {366, 417}

\bibitem[\protect\citeauthoryear{{Falceta-Gon{\c{c}}alves}, {Caproni},
  {Abraham}, {Teixeira}  \& {de Gouveia Dal Pino}}{{Falceta-Gon{\c{c}}alves}
  et~al.}{2010}]{Falceta-Goncalves2010}
{Falceta-Gon{\c{c}}alves} D.,  {Caproni} A.,  {Abraham} Z.,  {Teixeira} D.~M.,
   {de Gouveia Dal Pino} E.~M.,  2010, \mn@doi [\apjl]
  {10.1088/2041-8205/713/1/L74}, \href
  {https://ui.adsabs.harvard.edu/abs/2010ApJ...713L..74F} {713, L74}

\bibitem[\protect\citeauthoryear{{Forman} et~al.,}{{Forman}
  et~al.}{2007}]{Forman2007}
{Forman} W.,  et~al., 2007, \mn@doi [\apj] {10.1086/519480}, \href
  {https://ui.adsabs.harvard.edu/abs/2007ApJ...665.1057F} {665, 1057}

\bibitem[\protect\citeauthoryear{{Gendron-Marsolais}
  et~al.,}{{Gendron-Marsolais} et~al.}{2020}]{Gendron-Marsolais2020}
{Gendron-Marsolais} M.,  et~al., 2020, \mn@doi [\mnras]
  {10.1093/mnras/staa2003}, \href
  {https://ui.adsabs.harvard.edu/abs/2020MNRAS.499.5791G} {499, 5791}

\bibitem[\protect\citeauthoryear{{Ghizzardi}, {Rossetti}  \&
  {Molendi}}{{Ghizzardi} et~al.}{2010}]{Ghizzardi2010}
{Ghizzardi} S.,  {Rossetti} M.,   {Molendi} S.,  2010, \mn@doi [\aap]
  {10.1051/0004-6361/200912496}, \href
  {https://ui.adsabs.harvard.edu/abs/2010A&A...516A..32G} {516, A32}

\bibitem[\protect\citeauthoryear{{Hatch}, {Crawford}, {Fabian}  \&
  {Johnstone}}{{Hatch} et~al.}{2005}]{Hatch2005}
{Hatch} N.~A.,  {Crawford} C.~S.,  {Fabian} A.~C.,   {Johnstone} R.~M.,  2005,
  \mn@doi [\mnras] {10.1111/j.1365-2966.2005.08787.x}, \href
  {https://ui.adsabs.harvard.edu/abs/2005MNRAS.358..765H} {358, 765}

\bibitem[\protect\citeauthoryear{{Hitomi Collaboration}, {Aharonian},
  {Akamatsu}, {Akimoto}, {Allen}, {Anabuki}, {Angelini}  \& {et al.}}{{Hitomi
  Collaboration} et~al.}{2016}]{Hitomi2016}
{Hitomi Collaboration} {Aharonian} F.,  {Akamatsu} H.,  {Akimoto} F.,  {Allen}
  S.~W.,  {Anabuki} N.,  {Angelini} L.,   {et al.} 2016, \mn@doi [\nat]
  {10.1038/nature18627}, \href
  {https://ui.adsabs.harvard.edu/abs/2016Natur.535..117H} {535, 117}

\bibitem[\protect\citeauthoryear{{Hitomi Collaboration} et~al.,}{{Hitomi
  Collaboration} et~al.}{2018}]{Hitomi2018}
{Hitomi Collaboration} et~al., 2018, \mn@doi [\pasj] {10.1093/pasj/psx138},
  \href {https://ui.adsabs.harvard.edu/abs/2018PASJ...70....9H} {70, 9}

\bibitem[\protect\citeauthoryear{{Hodgson} et~al.,}{{Hodgson}
  et~al.}{2021}]{Hodgson2021}
{Hodgson} J.~A.,  et~al., 2021, arXiv e-prints, \href
  {https://ui.adsabs.harvard.edu/abs/2021arXiv210403081H} {p. arXiv:2104.03081}

\bibitem[\protect\citeauthoryear{{Lim}, {Ohyama}, {Chi-Hung}, {Dinh-V-Trung}
  \& {Shiang-Yu}}{{Lim} et~al.}{2012}]{Lim2012}
{Lim} J.,  {Ohyama} Y.,  {Chi-Hung} Y.,  {Dinh-V-Trung}  {Shiang-Yu} W.,  2012,
  \mn@doi [\apj] {10.1088/0004-637X/744/2/112}, \href
  {https://ui.adsabs.harvard.edu/abs/2012ApJ...744..112L} {744, 112}

\bibitem[\protect\citeauthoryear{{Loken}, {Roettiger}, {Burns}  \&
  {Norman}}{{Loken} et~al.}{1995}]{Loken1995}
{Loken} C.,  {Roettiger} K.,  {Burns} J.~O.,   {Norman} M.,  1995, \mn@doi
  [\apj] {10.1086/175674}, \href
  {https://ui.adsabs.harvard.edu/abs/1995ApJ...445...80L} {445, 80}

\bibitem[\protect\citeauthoryear{{Marinacci} et~al.,}{{Marinacci}
  et~al.}{2018}]{Marinacci2018}
{Marinacci} F.,  et~al., 2018, \mn@doi [\mnras] {10.1093/mnras/sty2206}, \href
  {https://ui.adsabs.harvard.edu/abs/2018MNRAS.480.5113M} {480, 5113}

\bibitem[\protect\citeauthoryear{{Mendygral}, {Jones}  \& {Dolag}}{{Mendygral}
  et~al.}{2012}]{Mendygral2012}
{Mendygral} P.~J.,  {Jones} T.~W.,   {Dolag} K.,  2012, \mn@doi [\apj]
  {10.1088/0004-637X/750/2/166}, \href
  {https://ui.adsabs.harvard.edu/abs/2012ApJ...750..166M} {750, 166}

\bibitem[\protect\citeauthoryear{{Nawaz}, {Bicknell}, {Wagner}, {Sutherland}
  \& {McNamara}}{{Nawaz} et~al.}{2016}]{Nawaz2016}
{Nawaz} M.~A.,  {Bicknell} G.~V.,  {Wagner} A.~Y.,  {Sutherland} R.~S.,
  {McNamara} B.~R.,  2016, \mn@doi [\mnras] {10.1093/mnras/stw330}, \href
  {https://ui.adsabs.harvard.edu/abs/2016MNRAS.458..802N} {458, 802}

\bibitem[\protect\citeauthoryear{{Nulsen}, {McNamara}, {Wise}  \&
  {David}}{{Nulsen} et~al.}{2005}]{Nulsen2005}
{Nulsen} P.~E.~J.,  {McNamara} B.~R.,  {Wise} M.~W.,   {David} L.~P.,  2005,
  \mn@doi [\apj] {10.1086/430845}, \href
  {https://ui.adsabs.harvard.edu/abs/2005ApJ...628..629N} {628, 629}

\bibitem[\protect\citeauthoryear{{Pakmor} \& {Springel}}{{Pakmor} \&
  {Springel}}{2013}]{Pakmor2013}
{Pakmor} R.,  {Springel} V.,  2013, \mn@doi [\mnras] {10.1093/mnras/stt428},
  \href {https://ui.adsabs.harvard.edu/abs/2013MNRAS.432..176P} {432, 176}

\bibitem[\protect\citeauthoryear{{Panagoulia}, {Fabian}, {Sanders}  \&
  {Hlavacek-Larrondo}}{{Panagoulia} et~al.}{2014}]{Panagoulia2014}
{Panagoulia} E.~K.,  {Fabian} A.~C.,  {Sanders} J.~S.,   {Hlavacek-Larrondo}
  J.,  2014, \mn@doi [\mnras] {10.1093/mnras/stu1499}, \href
  {https://ui.adsabs.harvard.edu/abs/2014MNRAS.444.1236P} {444, 1236}

\bibitem[\protect\citeauthoryear{{Pedlar}, {Ghataure}, {Davies}, {Harrison},
  {Perley}, {Crane}  \& {Unger}}{{Pedlar} et~al.}{1990}]{Pedlar1990}
{Pedlar} A.,  {Ghataure} H.~S.,  {Davies} R.~D.,  {Harrison} B.~A.,  {Perley}
  R.,  {Crane} P.~C.,   {Unger} S.~W.,  1990, \mnras, \href
  {https://ui.adsabs.harvard.edu/abs/1990MNRAS.246..477P} {246, 477}

\bibitem[\protect\citeauthoryear{{Randall} et~al.,}{{Randall}
  et~al.}{2011}]{Randall2011}
{Randall} S.~W.,  et~al., 2011, \mn@doi [\apj] {10.1088/0004-637X/726/2/86},
  \href {https://ui.adsabs.harvard.edu/abs/2011ApJ...726...86R} {726, 86}

\bibitem[\protect\citeauthoryear{{Randall} et~al.,}{{Randall}
  et~al.}{2015}]{Randall2015}
{Randall} S.~W.,  et~al., 2015, \mn@doi [\apj] {10.1088/0004-637X/805/2/112},
  \href {https://ui.adsabs.harvard.edu/abs/2015ApJ...805..112R} {805, 112}

\bibitem[\protect\citeauthoryear{{Roediger}, {Br{\"u}ggen}, {Simionescu},
  {B{\"o}hringer}, {Churazov}  \& {Forman}}{{Roediger}
  et~al.}{2011}]{Roediger2011}
{Roediger} E.,  {Br{\"u}ggen} M.,  {Simionescu} A.,  {B{\"o}hringer} H.,
  {Churazov} E.,   {Forman} W.~R.,  2011, \mn@doi [Mon.\ Not.\ R.\ Astron.\
  Soc.] {10.1111/j.1365-2966.2011.18279.x}, \href
  {http://adsabs.harvard.edu/abs/2011MNRAS.413.2057R} {413, 2057}

\bibitem[\protect\citeauthoryear{{Roediger}, {Lovisari}, {Dupke}, {Ghizzardi},
  {Br{\"u}ggen}, {Kraft}  \& {Machacek}}{{Roediger}
  et~al.}{2012}]{Roediger2012}
{Roediger} E.,  {Lovisari} L.,  {Dupke} R.,  {Ghizzardi} S.,  {Br{\"u}ggen} M.,
   {Kraft} R.~P.,   {Machacek} M.~E.,  2012, \mn@doi [Mon.\ Not.\ R.\ Astron.\
  Soc.] {10.1111/j.1365-2966.2011.20287.x}, \href
  {http://adsabs.harvard.edu/abs/2012MNRAS.420.3632R} {420, 3632}

\bibitem[\protect\citeauthoryear{{Salom{\'e}} et~al.,}{{Salom{\'e}}
  et~al.}{2006}]{Salome2006}
{Salom{\'e}} P.,  et~al., 2006, \mn@doi [\aap] {10.1051/0004-6361:20054745},
  \href {https://ui.adsabs.harvard.edu/abs/2006A&A...454..437S} {454, 437}

\bibitem[\protect\citeauthoryear{{Sanders}, {Fabian}  \& {Taylor}}{{Sanders}
  et~al.}{2009}]{Sanders2009}
{Sanders} J.~S.,  {Fabian} A.~C.,   {Taylor} G.~B.,  2009, \mn@doi [\mnras]
  {10.1111/j.1365-2966.2009.14892.x}, \href
  {https://ui.adsabs.harvard.edu/abs/2009MNRAS.396.1449S} {396, 1449}

\bibitem[\protect\citeauthoryear{{Sanders} et~al.,}{{Sanders}
  et~al.}{2020}]{Sanders2020}
{Sanders} J.~S.,  et~al., 2020, \mn@doi [\aap] {10.1051/0004-6361/201936468},
  \href {https://ui.adsabs.harvard.edu/abs/2020A&A...633A..42S} {633, A42}

\bibitem[\protect\citeauthoryear{{Schellenberger}, {David}, {Vrtilek},
  {O'Sullivan}, {Giacintucci}, {Forman}, {Jones}  \&
  {Venturi}}{{Schellenberger} et~al.}{2021}]{Schellenberger2021}
{Schellenberger} G.,  {David} L.~P.,  {Vrtilek} J.,  {O'Sullivan} E.,
  {Giacintucci} S.,  {Forman} W.,  {Jones} C.,   {Venturi} T.,  2021, \mn@doi
  [\apj] {10.3847/1538-4357/abc488}, \href
  {https://ui.adsabs.harvard.edu/abs/2021ApJ...906...16S} {906, 16}

\bibitem[\protect\citeauthoryear{{Soker} \& {Bisker}}{{Soker} \&
  {Bisker}}{2006}]{Soker2006}
{Soker} N.,  {Bisker} G.,  2006, \mn@doi [\mnras]
  {10.1111/j.1365-2966.2006.10313.x}, \href
  {https://ui.adsabs.harvard.edu/abs/2006MNRAS.369.1115S} {369, 1115}

\bibitem[\protect\citeauthoryear{{Springel}}{{Springel}}{2010}]{Springel2010}
{Springel} V.,  2010, \mn@doi [\mnras] {10.1111/j.1365-2966.2009.15715.x},
  \href {https://ui.adsabs.harvard.edu/abs/2010MNRAS.401..791S} {401, 791}

\bibitem[\protect\citeauthoryear{{Sternberg} \& {Soker}}{{Sternberg} \&
  {Soker}}{2008}]{SternbergSoker2008}
{Sternberg} A.,  {Soker} N.,  2008, \mn@doi [\mnras]
  {10.1111/j.1365-2966.2007.12802.x}, \href
  {https://ui.adsabs.harvard.edu/abs/2008MNRAS.384.1327S} {384, 1327}

\bibitem[\protect\citeauthoryear{{Tashiro}, {Maejima}, {Toda}, {Kelley},
  {Reichenthal}, {Lobell}, {Petre}  \& {et al.}}{{Tashiro}
  et~al.}{2018}]{XRISM2018}
{Tashiro} M.,  {Maejima} H.,  {Toda} K.,  {Kelley} R.,  {Reichenthal} L.,
  {Lobell} J.,  {Petre} R.,   {et al.} 2018, in {den Herder} J.-W.~A.,
  {Nikzad} S.,   {Nakazawa} K.,  eds,  Society of Photo-Optical Instrumentation
  Engineers (SPIE) Conference Series Vol. 10699, Space Telescopes and
  Instrumentation 2018: Ultraviolet to Gamma Ray. p. 1069922,
  \mn@doi{10.1117/12.2309455}

\bibitem[\protect\citeauthoryear{{Walker}, {Hlavacek-Larrondo},
  {Gendron-Marsolais}, {Fabian}, {Intema}, {Sanders}, {Bamford}  \& {van
  Weeren}}{{Walker} et~al.}{2017}]{Walker2017}
{Walker} S.~A.,  {Hlavacek-Larrondo} J.,  {Gendron-Marsolais} M.,  {Fabian}
  A.~C.,  {Intema} H.,  {Sanders} J.~S.,  {Bamford} J.~T.,   {van Weeren} R.,
  2017, \mn@doi [\mnras] {10.1093/mnras/stx640}, \href
  {https://ui.adsabs.harvard.edu/abs/2017MNRAS.468.2506W} {468, 2506}

\bibitem[\protect\citeauthoryear{{Weinberger}, {Ehlert}, {Pfrommer}, {Pakmor}
  \& {Springel}}{{Weinberger} et~al.}{2017}]{Weinberger2017}
{Weinberger} R.,  {Ehlert} K.,  {Pfrommer} C.,  {Pakmor} R.,   {Springel} V.,
  2017, \mn@doi [\mnras] {10.1093/mnras/stx1409}, \href
  {https://ui.adsabs.harvard.edu/abs/2017MNRAS.470.4530W} {470, 4530}

\bibitem[\protect\citeauthoryear{{Wise}, {McNamara}, {Nulsen}, {Houck}  \&
  {David}}{{Wise} et~al.}{2007}]{Wise2007}
{Wise} M.~W.,  {McNamara} B.~R.,  {Nulsen} P.~E.~J.,  {Houck} J.~C.,   {David}
  L.~P.,  2007, \mn@doi [\apj] {10.1086/512767}, \href
  {https://ui.adsabs.harvard.edu/abs/2007ApJ...659.1153W} {659, 1153}

\bibitem[\protect\citeauthoryear{{ZuHone} \& {Roediger}}{{ZuHone} \&
  {Roediger}}{2016}]{ZR2016}
{ZuHone} J.~A.,  {Roediger} E.,  2016, \mn@doi [Journal of Plasma Physics]
  {10.1017/S0022377816000544}, \href
  {https://ui.adsabs.harvard.edu/abs/2016JPlPh..82c5301Z} {82, 535820301}

\bibitem[\protect\citeauthoryear{{ZuHone}, {Markevitch}  \& {Johnson}}{{ZuHone}
  et~al.}{2010}]{ZuHone2010}
{ZuHone} J.~A.,  {Markevitch} M.,   {Johnson} R.~E.,  2010, \mn@doi [\apj]
  {10.1088/0004-637X/717/2/908}, \href
  {https://ui.adsabs.harvard.edu/abs/2010ApJ...717..908Z} {717, 908}

\bibitem[\protect\citeauthoryear{{ZuHone}, {Miller}, {Simionescu}  \&
  {Bautz}}{{ZuHone} et~al.}{2016}]{ZuHone2016}
{ZuHone} J.~A.,  {Miller} E.~D.,  {Simionescu} A.,   {Bautz} M.~W.,  2016,
  \mn@doi [\apj] {10.3847/0004-637X/821/1/6}, \href
  {https://ui.adsabs.harvard.edu/abs/2016ApJ...821....6Z} {821, 6}

\bibitem[\protect\citeauthoryear{{ZuHone}, {Miller}, {Bulbul}  \&
  {Zhuravleva}}{{ZuHone} et~al.}{2018}]{ZuHone2018}
{ZuHone} J.~A.,  {Miller} E.~D.,  {Bulbul} E.,   {Zhuravleva} I.,  2018,
  \mn@doi [\apj] {10.3847/1538-4357/aaa4b3}, \href
  {https://ui.adsabs.harvard.edu/abs/2018ApJ...853..180Z} {853, 180}

\bibitem[\protect\citeauthoryear{{ZuHone}, {Zavala}  \&
  {Vogelsberger}}{{ZuHone} et~al.}{2019}]{ZuHone2019}
{ZuHone} J.~A.,  {Zavala} J.,   {Vogelsberger} M.,  2019, \mn@doi [\apj]
  {10.3847/1538-4357/ab321d}, \href
  {https://ui.adsabs.harvard.edu/abs/2019ApJ...882..119Z} {882, 119}

\bibitem[\protect\citeauthoryear{{ZuHone}, {Markevitch}, {Weinberger}, {Nulsen}
   \& {Ehlert}}{{ZuHone} et~al.}{2020}]{ZuHone2020}
{ZuHone} J.~A.,  {Markevitch} M.,  {Weinberger} R.,  {Nulsen} P.,   {Ehlert}
  K.,  2020, arXiv e-prints, \href
  {https://ui.adsabs.harvard.edu/abs/2020arXiv201202001Z} {p. arXiv:2012.02001}

\makeatother
\end{thebibliography}




\appendix

\section{Supplemental X-ray Surface Brightness and Cosmic Ray Energy Density Maps}\label{sec:supp_plots}

\begin{figure*}
\begin{center}
\includegraphics[width=0.95\textwidth]{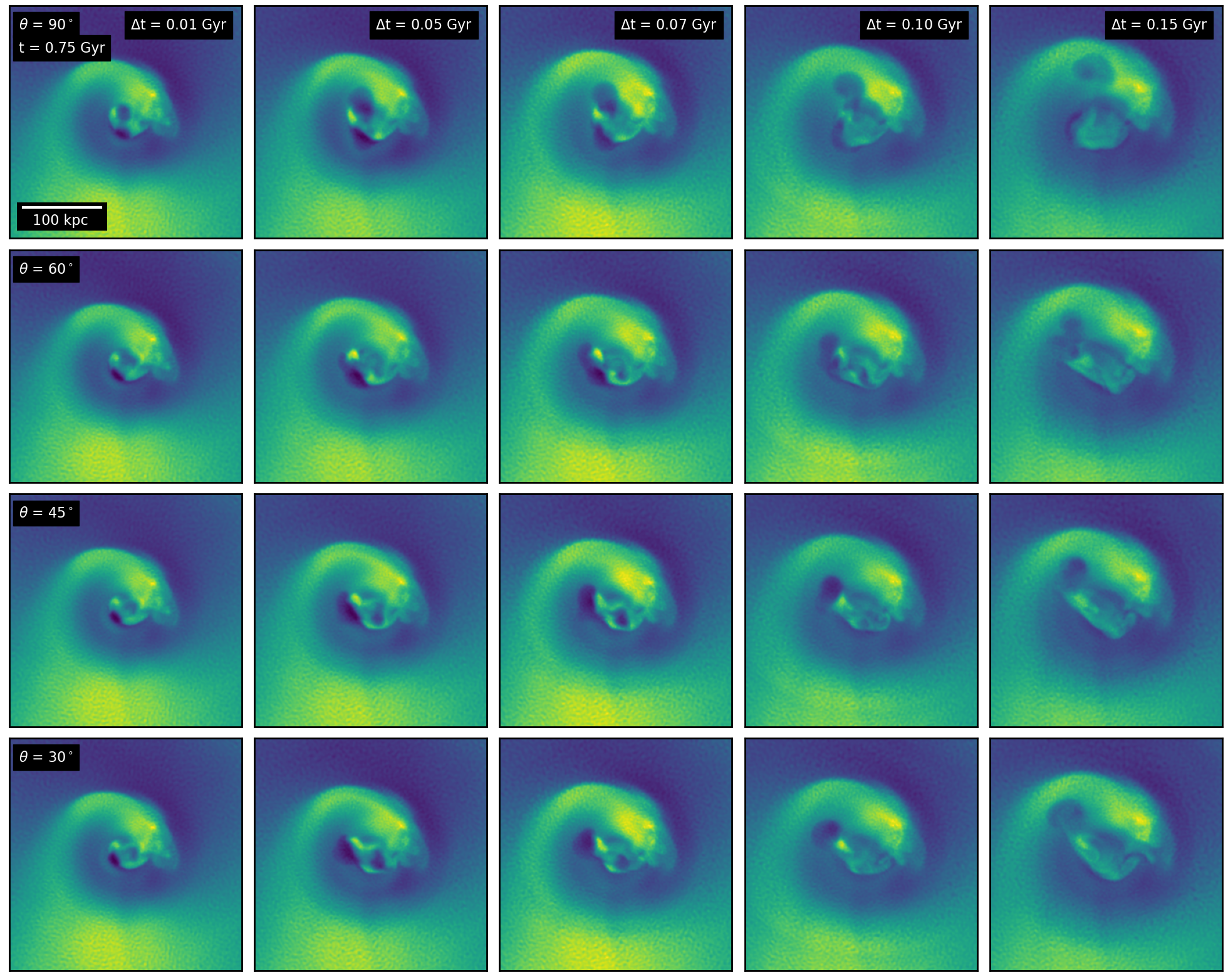}
\caption{X-ray surface brightness residuals of simulation with sloshing gas
		 motions and AGN jets, where the jet axis is 90, 60, 45, and 30
		 degrees from the horizontal. Epochs shown are 0.01, 0.05, 0.07,
		 0.1, and 0.15~Gyr from the jet ignition.\label{fig:sx_210_1}}
\end{center}
\end{figure*}

\begin{figure*}
\centering
\includegraphics[width=0.95\textwidth]{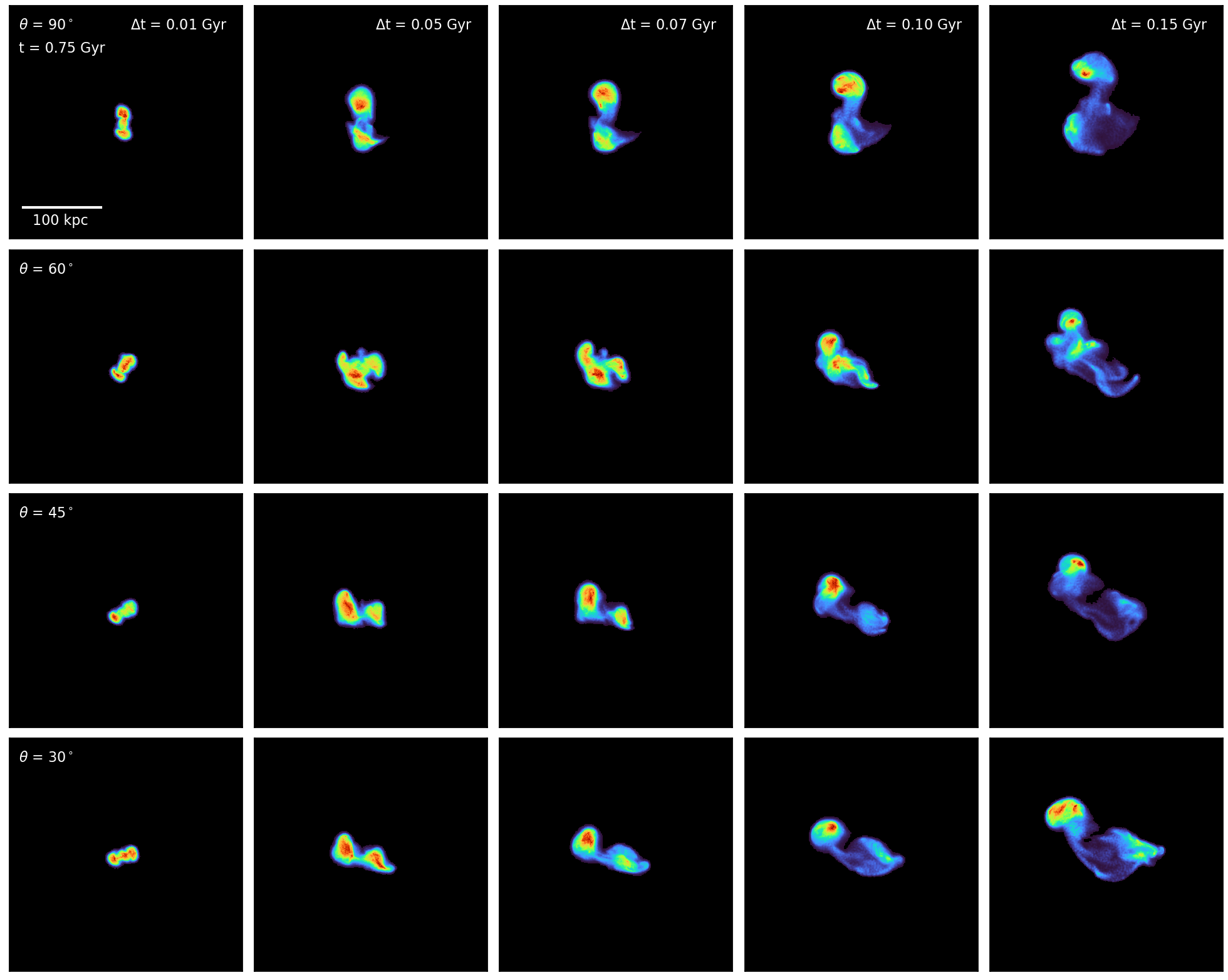}
\caption{Projected CR energy density of simulation with sloshing gas
		 motions and AGN jets, where the jet axis is 90, 60, 45, and 30
		 degrees from the horizontal. Epochs shown are $\Delta{t}$ = 0.01, 0.05, 0.07, 0.1, and 0.15~Gyr from the jet ignition at $t$ = 0.75~Gyr. \label{fig:cr_210_1}}
\end{figure*}
		
\begin{figure*}
\centering
\includegraphics[width=0.95\textwidth]{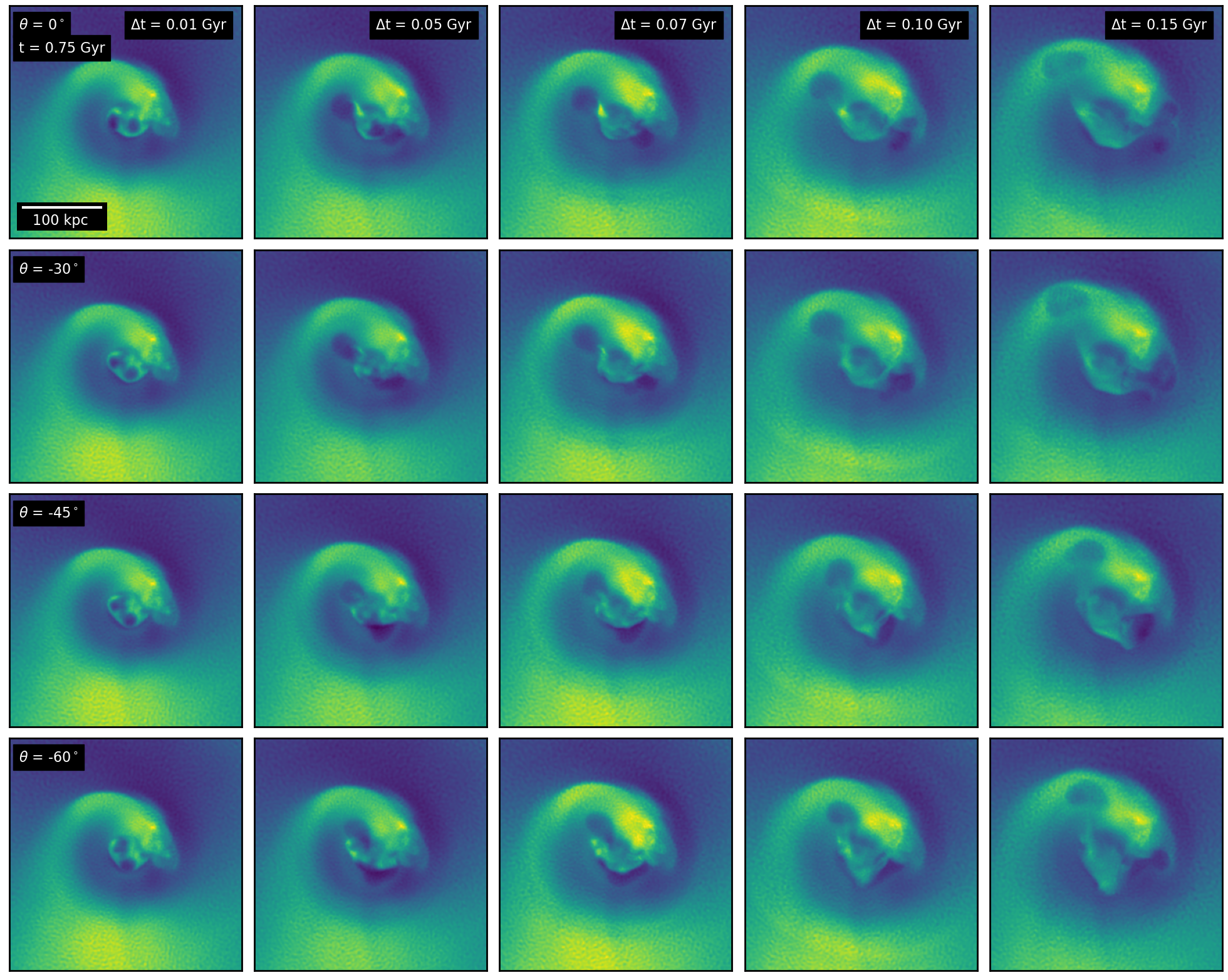}
\caption{X-ray surface brightness residuals of simulation with sloshing gas
			motions and AGN jets, where the jet axis is 0, -30, -45, and -60
			degrees from the horizontal. Epochs shown are $\Delta{t}$ = 0.01, 0.05, 0.07, 0.1, and 0.15~Gyr from the jet ignition at $t$ = 0.75~Gyr.\label{fig:sx_210_2}}
\end{figure*}

\begin{figure*}
\centering
\includegraphics[width=0.95\textwidth]{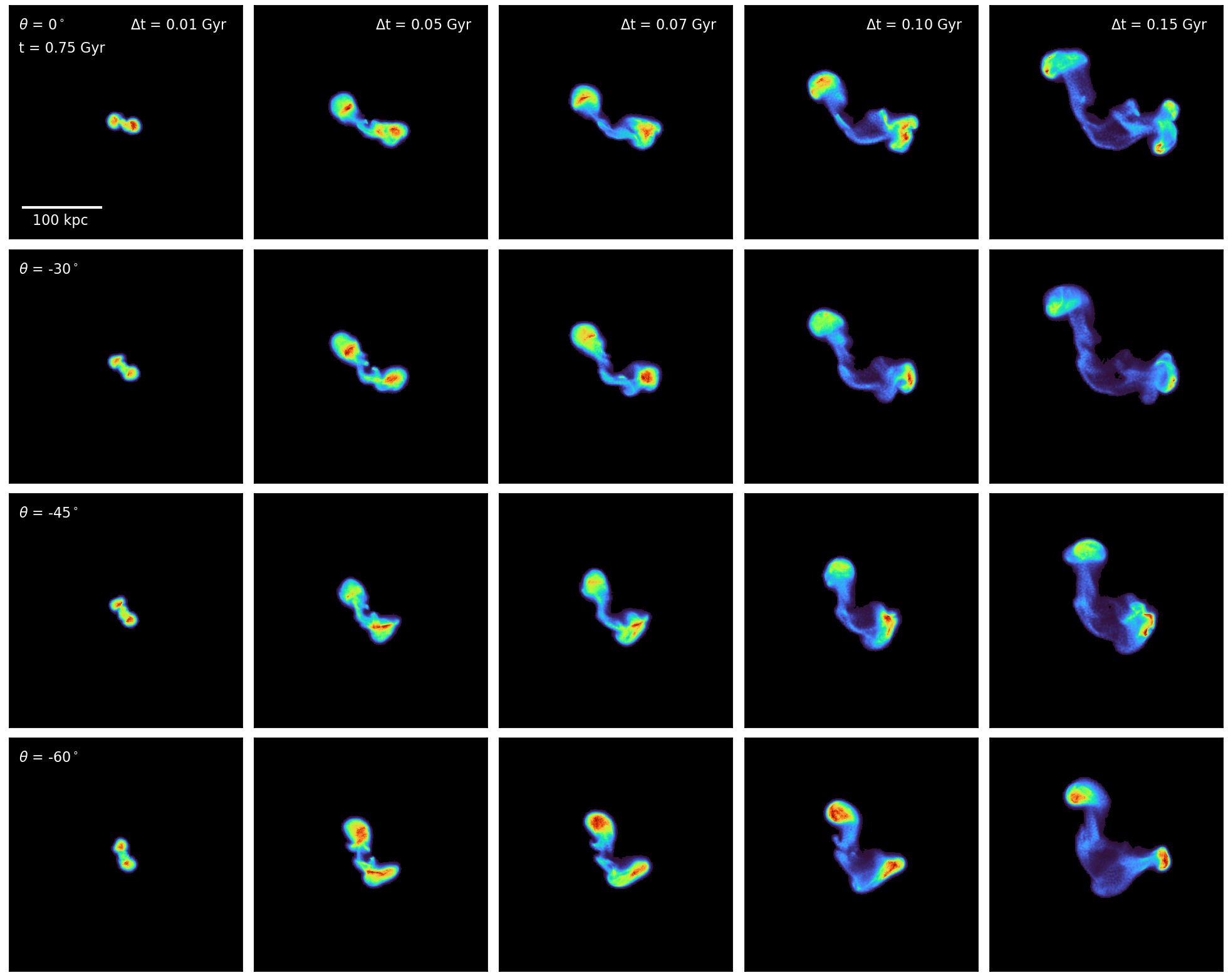}
\caption{Projected CR energy density of simulation with sloshing gas
			motions and AGN jets, where the jet axis is 0, -30, -45, and -60
			degrees from the horizontal. Epochs shown are $\Delta{t}$ = 0.01, 0.05, 0.07,
			0.1, and 0.15~Gyr from the jet ignition at $t$ = 0.75~Gyr.\label{fig:cr_210_2}}
\end{figure*}

\begin{figure*}
\begin{center}
\includegraphics[width=0.95\textwidth]{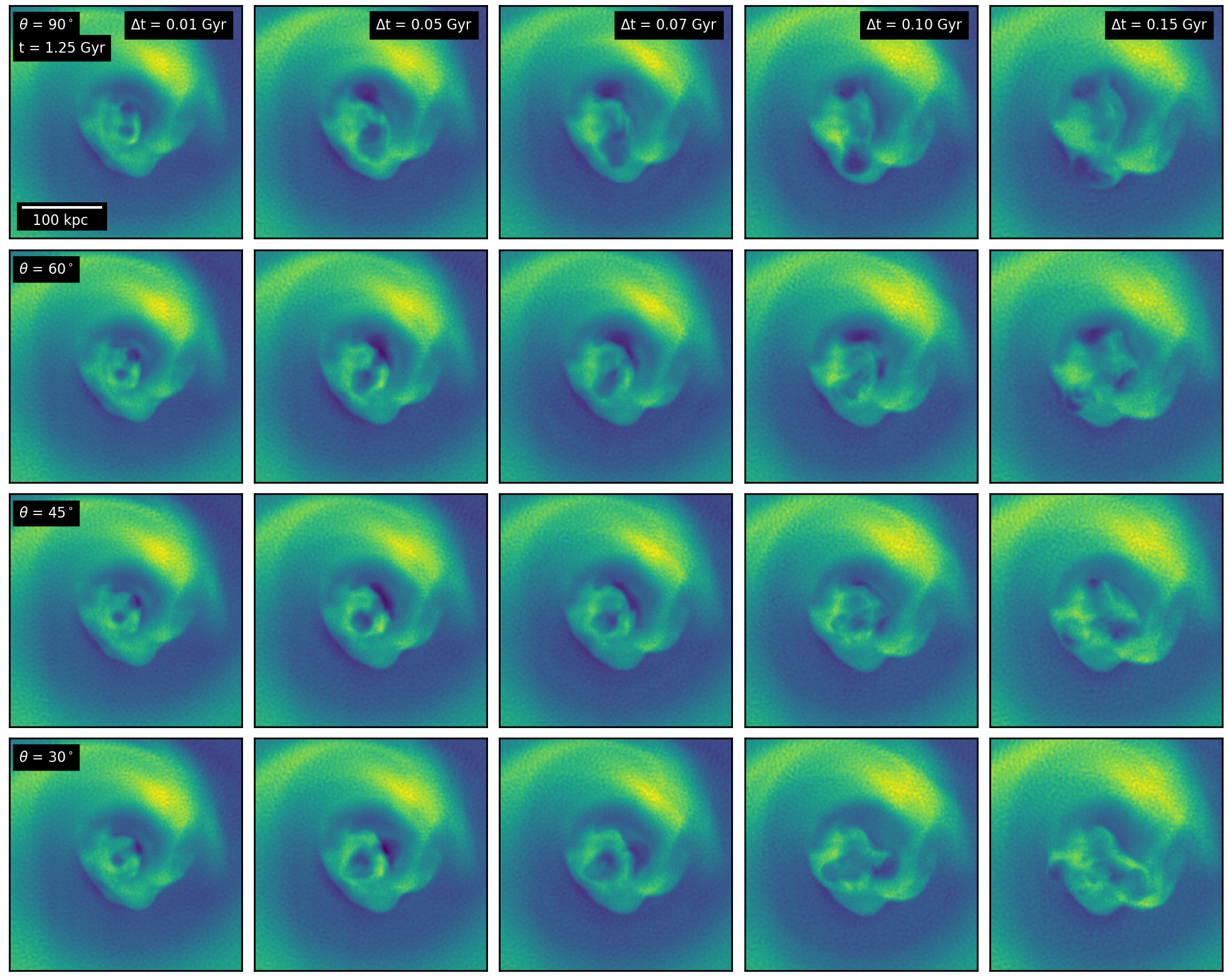}
\caption{X-ray surface brightness residuals of simulation with sloshing gas
			motions and AGN jets, where the jet axis is 90, 60, 45, and 30
			degrees from the horizontal. Epochs shown are $\Delta{t}$ = 0.01, 0.05, 0.07,
			0.1, and 0.15~Gyr from the jet ignition at $t$ = 1.25~Gyr.\label{fig:sx_260_1}}
\end{center}
\end{figure*}

\begin{figure*}
\centering
\includegraphics[width=0.95\textwidth]{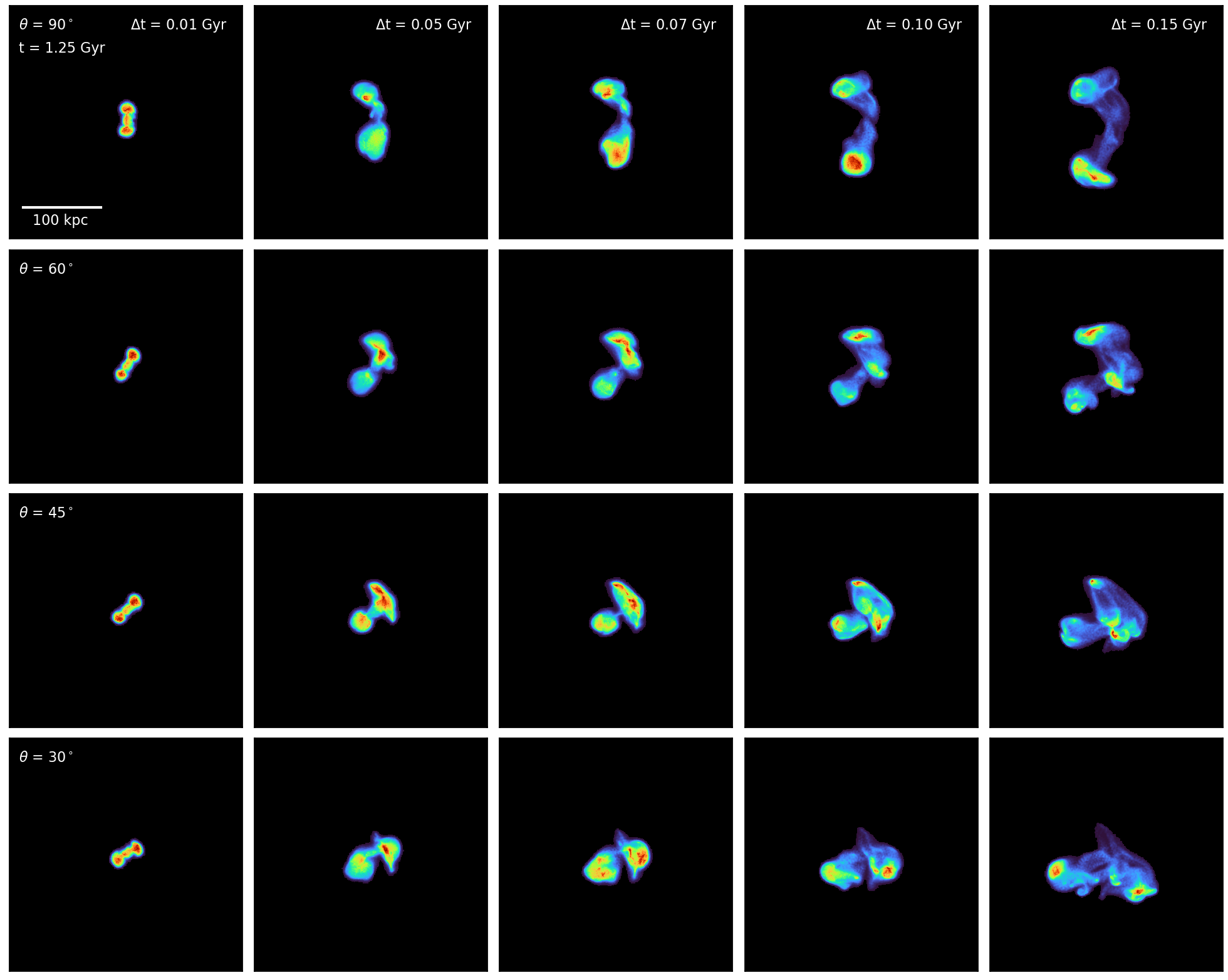}
\caption{Projected CR energy density of simulation with sloshing gas
			motions and AGN jets, where the jet axis is 90, 60, 45, and 30
			degrees from the horizontal. Epochs shown are $\Delta{t}$ = 0.01, 0.05, 0.07,
			0.1, and 0.15~Gyr from the jet ignition at $t$ = 1.25~Gyr.\label{fig:cr_260_1}}
\end{figure*}

\begin{figure*}
\centering
\includegraphics[width=0.95\textwidth]{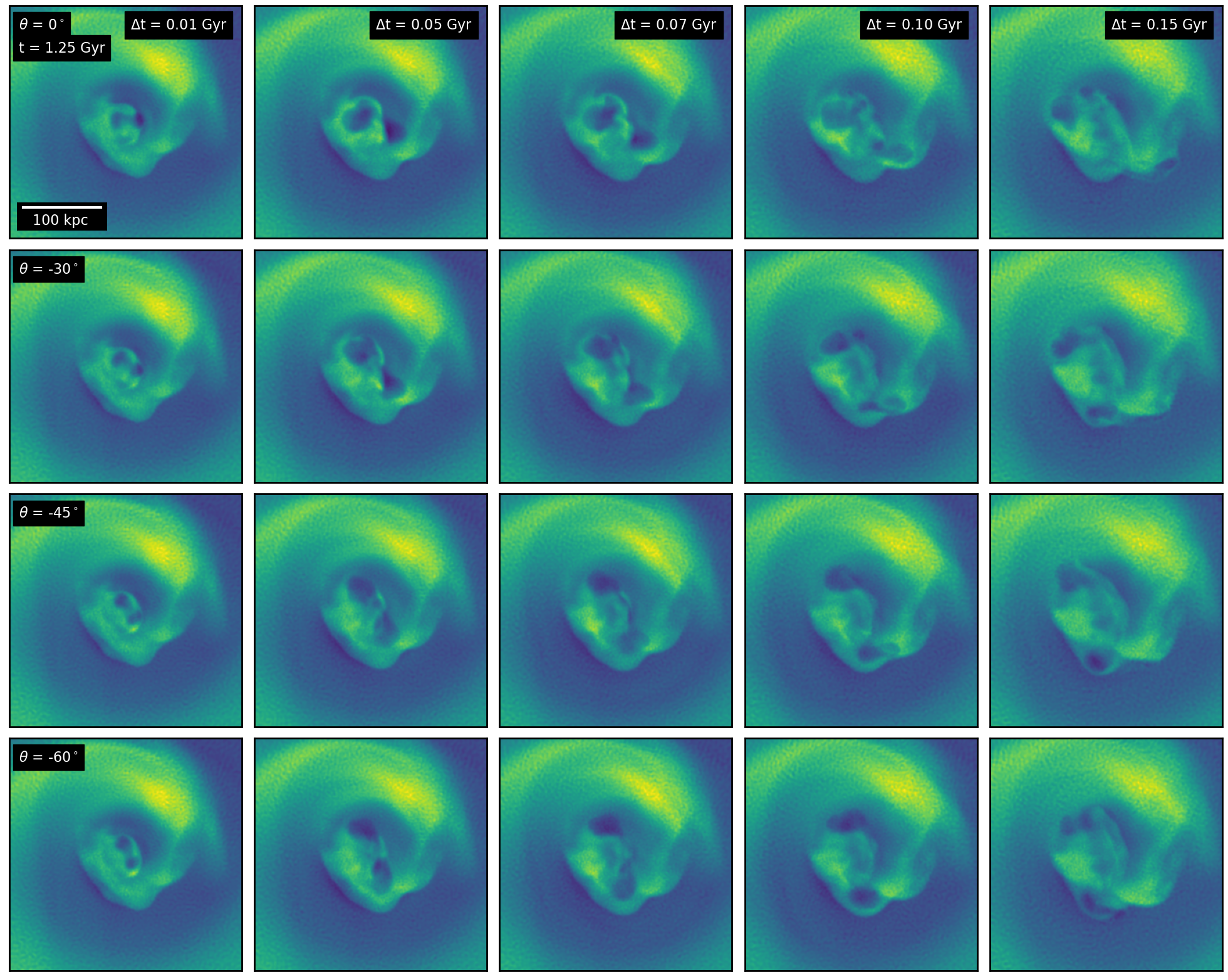}
\caption{X-ray surface brightness residuals of simulation with sloshing gas
		 motions and AGN jets, where the jet axis is 0, -30, -45, and -60
		 degrees from the horizontal. Epochs shown are $\Delta{t}$ = 0.01, 0.05, 0.07,
		 0.1, and 0.15~Gyr from the jet ignition at $t$ = 1.25~Gyr.\label{fig:sx_260_2}}
\end{figure*}

\begin{figure*}
\centering
\includegraphics[width=0.95\textwidth]{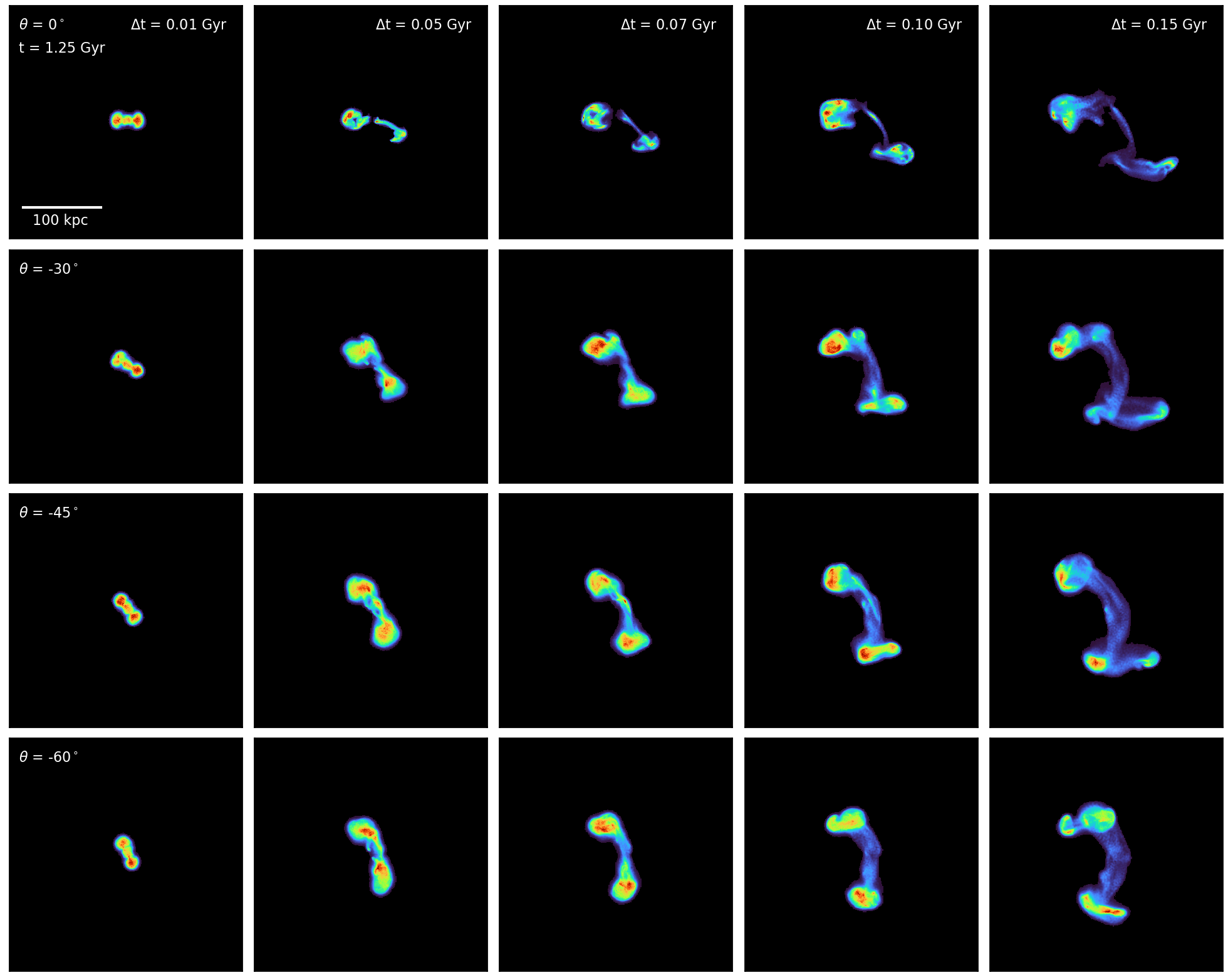}
\caption{Projected CR energy density of simulation with sloshing gas
		 motions and AGN jets, where the jet axis is 0, -30, -45, and -60
		 degrees from the horizontal. Epochs shown are $\Delta{t}$ = 0.01, 0.05, 0.07,
		 0.1, and 0.15~Gyr from the jet ignition at $t$ = 1.25~Gyr.\label{fig:cr_260_2}}
\end{figure*}

\begin{figure*}
\begin{center}
\includegraphics[width=0.95\textwidth]{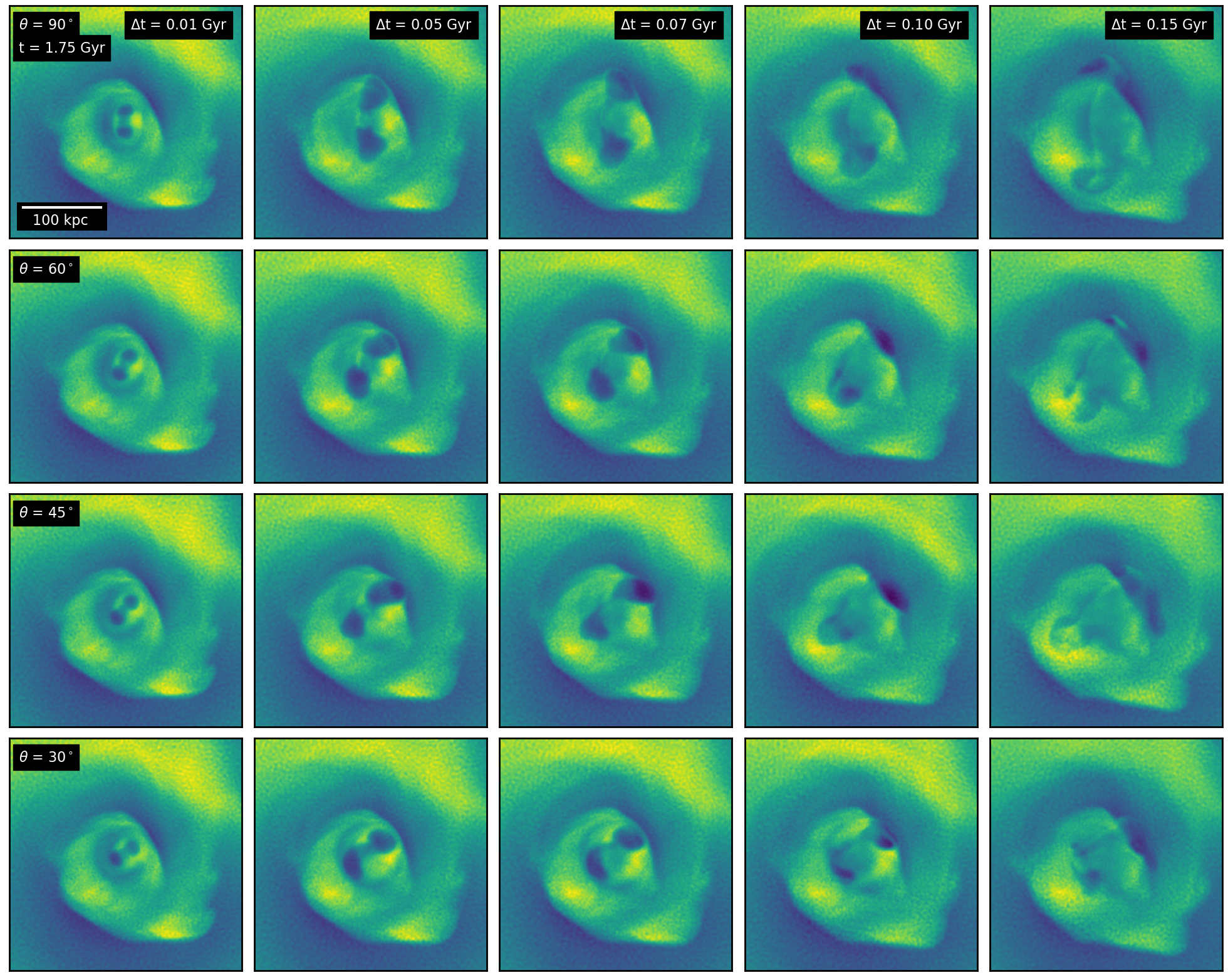}
\caption{X-ray surface brightness residuals of simulation with sloshing gas
			motions and AGN jets, where the jet axis is 90, 60, 45, and 30
			degrees from the horizontal. Epochs shown are $\Delta{t}$ = 0.01, 0.05, 0.07,
			0.1, and 0.15~Gyr from the jet ignition at $t$ = 1.75~Gyr.\label{fig:sx_310_1}}
\end{center}
\end{figure*}

\begin{figure*}
\centering
\includegraphics[width=0.95\textwidth]{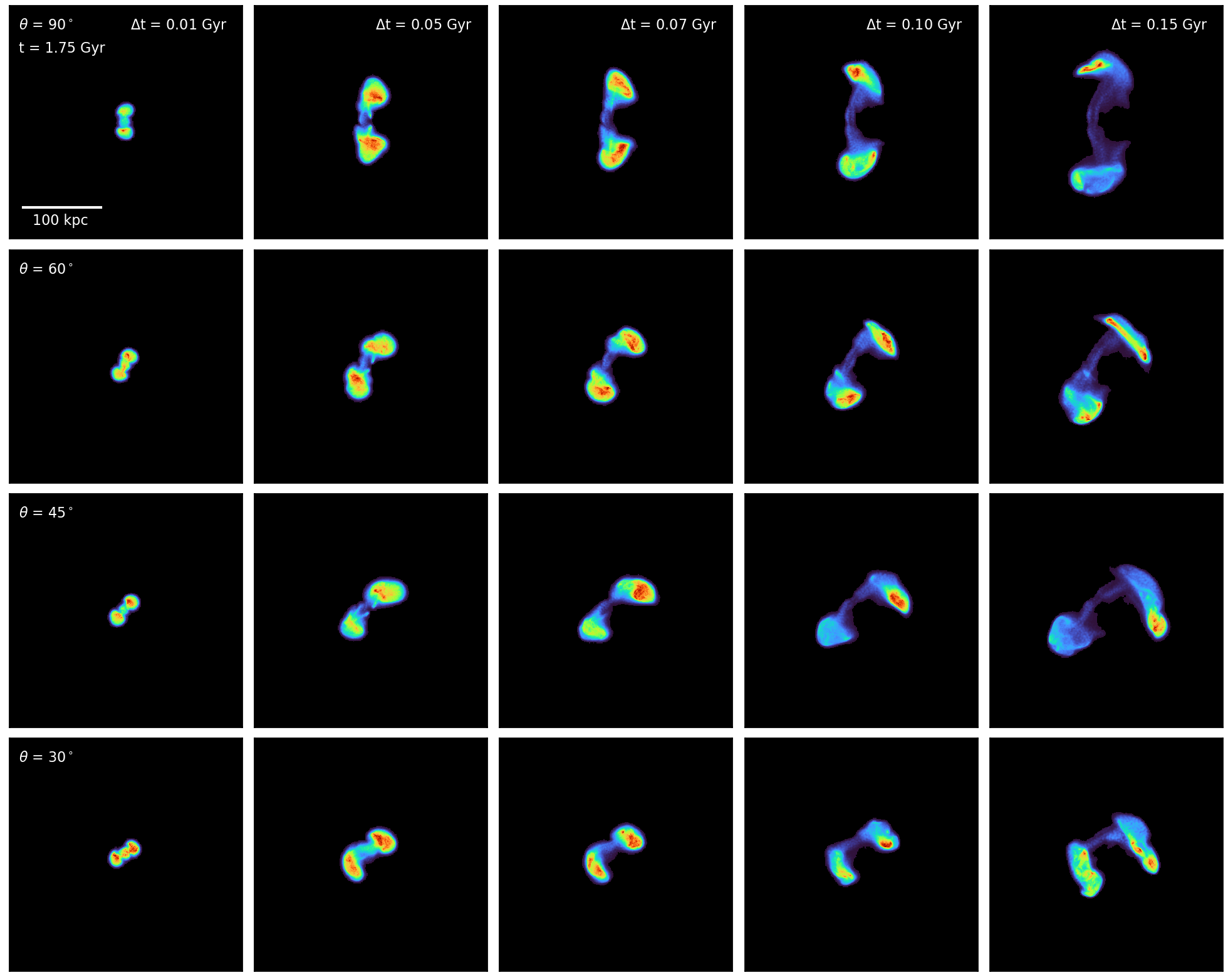}
\caption{Projected CR energy density of simulation with sloshing gas
			motions and AGN jets, where the jet axis is 90, 60, 45, and 30
			degrees from the horizontal. Epochs shown are $\Delta{t}$ = 0.01, 0.05, 0.07,
			0.1, and 0.15~Gyr from the jet ignition at $t$ = 1.75~Gyr.\label{fig:cr_310_1}}
\end{figure*}

\begin{figure*}
\centering
\includegraphics[width=0.95\textwidth]{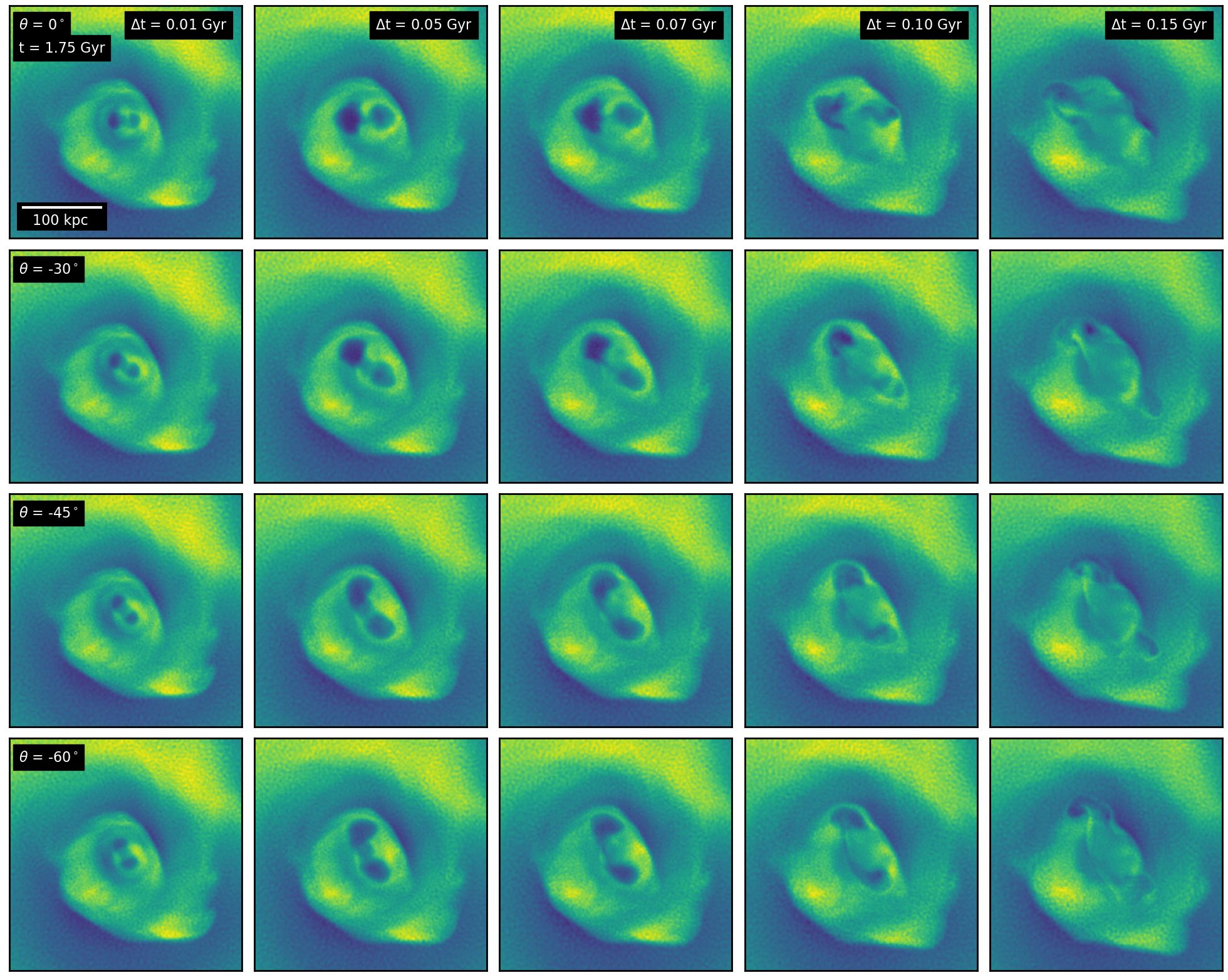}
\caption{X-ray surface brightness residuals of simulation with sloshing gas
			motions and AGN jets, where the jet axis is 0, -30, -45, and -60
			degrees from the horizontal. Epochs shown are $\Delta{t}$ = 0.01, 0.05, 0.07,
			0.1, and 0.15~Gyr from the jet ignition at $t$ = 1.75~Gyr.\label{fig:sx_310_2}}
\end{figure*}

\begin{figure*}
\centering
\includegraphics[width=0.95\textwidth]{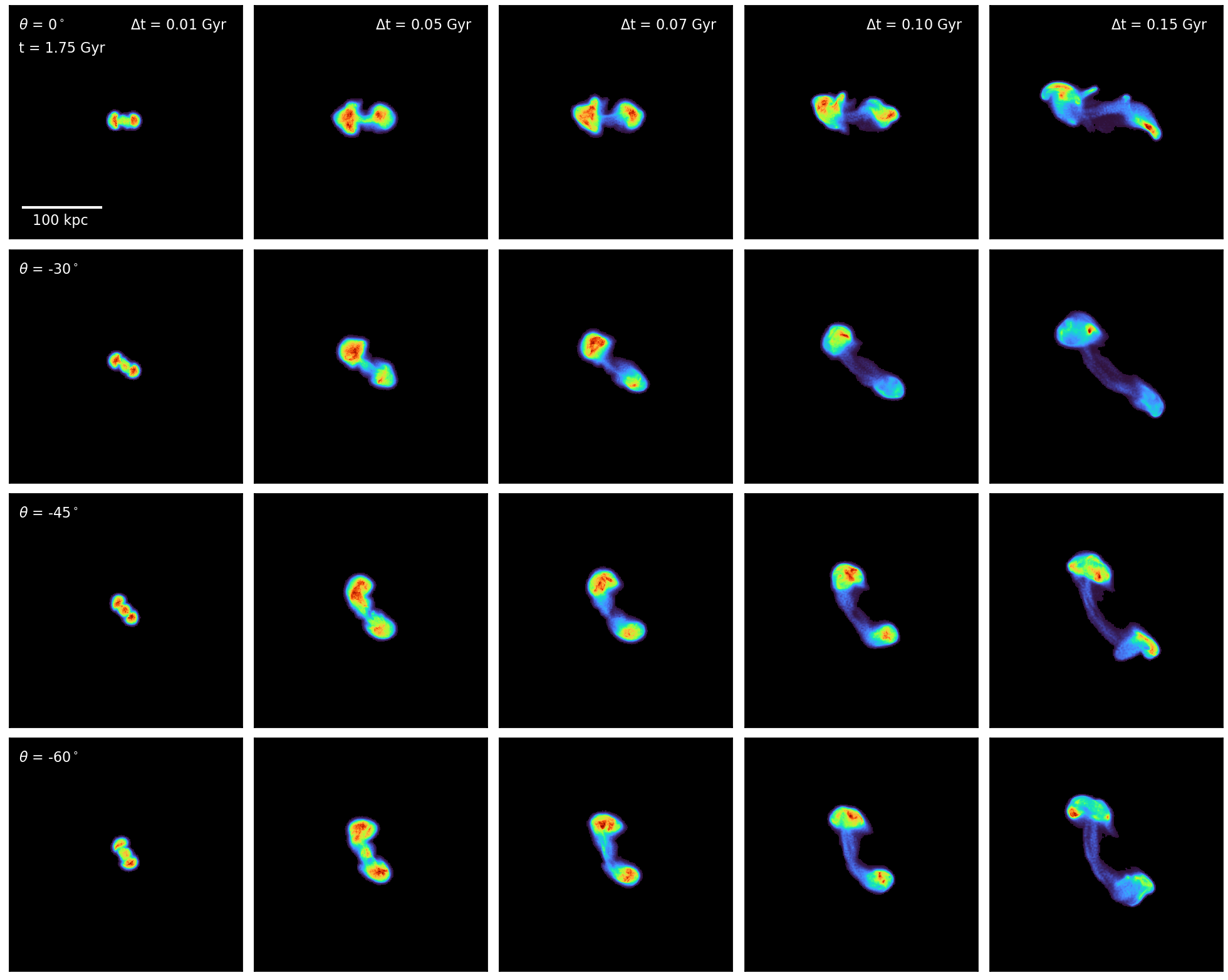}
\caption{Projected CR energy density of simulation with sloshing gas
			motions and AGN jets, where the jet axis is 0, -30, -45, and -60
			degrees from the horizontal. Epochs shown are $\Delta{t}$ = 0.01, 0.05, 0.07,
			0.1, and 0.15~Gyr from the jet ignition at $t$ = 1.75~Gyr.\label{fig:cr_310_2}}
\end{figure*}
	

\bsp	
\label{lastpage}
\end{document}